\documentclass[a4paper,12pt]{article}
\usepackage{amssymb}
\usepackage{amsmath}
\usepackage{epsfig}
\usepackage{subfig}
\usepackage{caption}
\usepackage{graphicx}
\usepackage{tikz}
\usepackage{latexsym}
\usepackage{rotating}
\usepackage{titlesec}
\newcommand{\squeezeup}{\vspace{-11mm}}
\usetikzlibrary{matrix}
\newtheorem{thm}{Theorem}[section]

\newtheorem{rem}[thm]{Remark}

\usepackage{relsize}
\title{Shifted nonlocal Kundu type equations: Soliton solutions}

\author{Asl{\i} Pekcan \thanks{aslipekcan@hacettepe.edu.tr} \\
{\small Department of Mathematics, Faculty of Science} \\
{\small Hacettepe University, 06800 Ankara - Turkey}
}

\date{\nonumber}
\begin{document}
\maketitle
\date{\nonumber}
\begin{abstract}
We study the local and shifted nonlocal reductions of the integrable coupled Kundu type system. We then consider particular cases of this system; namely Chen-Lee-Liu, Gerdjikov-Ivanov, and Kaup-Newell systems. We obtain one- and two-soliton solutions of these systems and their local and shifted nonlocal reductions by the Hirota bilinear method. We present particular examples for one- and two-soliton solutions of the reduced local and shifted nonlocal Chen-Lee-Liu, Gerdjikov-Ivanov, and Kaup-Newell equations. \\

\end{abstract}

\section{Introduction}

The integrable coupled Kundu type system \cite{Kakei}-\cite{TD} is
\begin{align}
&aq_t-iq_{xx}+(4\beta+1)q^2r_x+4\beta qq_xr-\frac{i}{2}(1+2\beta)(4\beta+1)q^3r^2=0,\label{cKundu-a}\\
&ar_t+ir_{xx}+(4\beta+1)r^2q_x+4\beta rr_xq+\frac{i}{2}(1+2\beta)(4\beta+1)q^2r^3=0,\label{cKundu-b}
\end{align}
where $a, \beta$ are constants. This system and its reductions have applications in plasma physics, quantum physics, and nonlinear fluids. By the local reduction $r(x,t)=\rho \bar{q}(x,t)$, $\rho \in \mathbb{R}$, where the bar notation stands for complex conjugation, we have the Kundu type equation
\begin{align}
&aq_t(x,t)-iq_{xx}(x,t)+(4\beta+1)\rho q^2(x,t)\bar{q}_x(x,t)+4\beta \rho \bar{q}(x,t)q(x,t)q_x(x,t)\nonumber\\
&-\frac{i}{2}(1+2\beta)(4\beta+1)\rho^2q^3(x,t)\bar{q}^2(x,t)=0,
\end{align}
where $a=\bar{a}$. This equation was first introduced by Kundu \cite{Kundu1}, \cite{Kundu2}. In \cite{Clarkson}, Clarkson and Cosgrove applied the Painlev\'{e} test to this equation for checking integrability. $N$-soliton solutions of the Kundu type equation were obtained via Riemann-Hilbert approach in \cite{WenZFan}.\\

For particular values of $\beta$, the system (\ref{cKundu-a}) and (\ref{cKundu-b}) is related to very well-known three types of coupled derivative nonlinear Schr\"{o}dinger (NLS) systems. If $\beta=-\frac{1}{4}$, then it reduces to the coupled Chen-Lee-Liu (CLL) system,
\begin{align}
&aq_t-iq_{xx}-qq_xr=0,\label{CLL-a}\\
&ar_t+ir_{xx}-rr_xq=0.\label{CLL-b}
\end{align}
The above system gives CLL equation \cite{CLL1}-\cite{CLL5} under the reduction $r(x,t)=\rho \bar{q}(x,t)$. When $\beta=0$, we have the coupled Gerdjikov-Ivanov (GI) system,
\begin{align}
&aq_t-iq_{xx}+q^2r_x-\frac{i}{2}q^3r^2=0,\label{GI-a}\\
&ar_t+ir_{xx}+r^2q_x+\frac{i}{2}q^2r^3=0.\label{GI-b}
\end{align}
Applying the reduction $r(x,t)=\rho \bar{q}(x,t)$ to this system yields GI equation \cite{GI1}-\cite{GI7}.

For $\beta=-\frac{1}{2}$, we get the coupled Kaup-Newell (KN) system,
\begin{align}
&aq_t-iq_{xx}-q^2r_x-2qq_xr=0,\label{KN-a}\\
&ar_t+ir_{xx}-r^2q_x-2rr_xq=0,\label{KN-b}
\end{align}
reducing to KN equation \cite{KN1}-\cite{KN6} by $r(x,t)=\rho \bar{q}(x,t)$. There are gauge transformations among the local
reduced CLL, GI, and KN equations but it is hard to obtain a solution of one equation from another one due to the complicated calculations \cite{Wadati}. Therefore one needs to work with these equations separately.

Nowadays, the notion of integrable nonlocal reductions is a hot topic in the theory of integrable systems. This notion was first introduced by Ablowitz and Musslimani \cite{AbMu1}-\cite{AbMu3}. The nonlocal reductions are actually special cases of discrete symmetry transformations \cite{origin}. If the nonlocal reductions are
applied consistently to an integrable system then one obtains space reversal, time reversal,
and space-time reversal nonlocal equations, which are also integrable. There are many nonlocal integrable equations e.g., nonlocal NLS equations \cite{AbMu1}-\cite{AbMu3}, \cite{chen}-\cite{Ma1}, nonlocal modified Korteweg-de Vries (mKdV) equations \cite{AbMu2}-\cite{chen}, \cite{GurPek3}, \cite{GurPek2}-\cite{Yan}, nonlocal GI equation \cite{LZYY}, \cite{ZD}, and so on \cite{Gurses}-\cite{GurPek5}.

Ablowitz and Musslimani \cite{AbMu4} have recently introduced new nonlocal reductions called shifted nonlocal reductions given by $r(x,t)=k \bar{q}(x_{0}-x,t)$, $r(x,t)=k q(x,-t+t_{0})$, $r(x,t)=k q(x_{0}-x,t_{0}-t)$, and $r(x,t)=k \bar{q}(x_{0}-x,t_{0}-t)$ where $x_{0}$ and $t_{0}$ are arbitrary real constants. By these shifted nonlocal reductions the NLS and MKdV systems reduce to the shifted nonlocal NLS and MKdV equations, respectively. When $x_0=t_0=0$ the shifted nonlocal equations reduce back to standard unshifted nonlocal forms of them. We have also studied shifted nonlocal NLS and MKdV equations and found one- and two-soliton solutions of these equations by the Hirota method in \cite{GurPek7}. In \cite{Dajun}, the authors considered space-time shifted nonlocal NLS and mKdV
hierarchies and find their solutions by double Wronskians. In this work we obtain new integrable shifted nonlocal reduced equations, namely integrable shifted nonlocal CLL, GI, and KN equations, from the integrable coupled Kundu type system.

In Section 2, we use two different bilinearizing transformations and get Hirota bilinear forms of CLL, GI, and KN systems which are particular cases of the Kundu type system. We obtain one- and two-soliton solutions of these systems. In Section 3, we give all consistent local and nonlocal shifted reductions of the Kundu type system for any value of the free parameter $\beta$. In Section 4, we present one- and two-soliton solutions of the reduced local and shifted nonlocal CLL, GI, and KN equations with some particular examples.

\section{Hirota bilinear method}

Here the bilinearizing transformations that we use
give the Hirota bilinear forms for particular values of the parameter $\beta$ in the Kundu type system (\ref{cKundu-a}) and (\ref{cKundu-b}).

Let
\begin{equation}\label{biltranCLLGI}
q=\frac{g}{f},\quad r=\frac{h}{s},
\end{equation}
in (\ref{cKundu-a}) and (\ref{cKundu-b}). We get the Hirota bilinear form as
\begin{align}
&(aD_t-iD_x^2)\{g\cdot f\}=0,\label{HirotaGIandCLL-a}\\
&(aD_t+iD_x^2)\{h\cdot s\}=0,\label{HirotaGIandCLL-b}\\
&D_x^2\{f\cdot s\}+\frac{i}{2}D_x\{g\cdot h\}=0,\label{HirotaGIandCLL-c}\\
&D_x\{f\cdot s\}-\frac{(8\beta+1)i}{2}gh=0,\label{HirotaGIandCLL-d}
\end{align}
which is only valid for $\beta=-\frac{1}{4}$ corresponding to
CLL system (\ref{CLL-a}) and (\ref{CLL-b}), and for $\beta=0$ corresponding to GI system (\ref{GI-a}) and (\ref{GI-b}). Soliton solutions of the GI system were also studied in \cite{ZGC} by Hirota method. But the Hirota bilinear form of the system and so the solutions obtained in \cite{ZGC} were given incorrectly.

If we insert
\begin{equation}\label{biltranKN}
q=\frac{gs}{f^2},\quad r=\frac{hf}{s^2}
\end{equation}
into the system (\ref{cKundu-a}) and (\ref{cKundu-b}), we obtain the Hirota bilinear form for KN system (\ref{KN-a}) and (\ref{KN-b}) i.e. for $\beta=-\frac{1}{2}$ as
\begin{align}
&(aD_t-iD_x^2)\{g\cdot f\}=0,\label{HirotaKN-a}\\
&(aD_t+iD_x^2)\{h\cdot s\}=0,\label{HirotaKN-b}\\
&(aD_t-iD_x^2)\{f\cdot s\}=0,\label{HirotaKN-c}\\
&D_x\{f\cdot s\}+\frac{i}{2}gh=0.\label{HirotaKN-d}
\end{align}
Here $D$ is the Hirota $D$-operator \cite{Hirota1} given by
\begin{equation}
D_t^nD_x^m\{F\cdot G\}=\Big(\frac{\partial}{\partial t}-\frac{\partial}{\partial t'}\Big)^n\Big(\frac{\partial}{\partial x}-\frac{\partial}{\partial x'}\Big)^m\,F(x,y,t)G(x',y',t')|_{x'=x,t'=t}.
\end{equation}
 We expand the functions $g$, $h$, $f$, and $s$ in the form of power series as
 \begin{align}\displaystyle
    & g=\sum_{j=1}^{\infty}g^{(2j-1)}\epsilon^{2j-1}, \quad h=\sum_{j=1}^{\infty}h^{(2j-1)}\epsilon^{2j-1},\nonumber\\
    & f=1+\sum_{j=1}^{\infty}f^{(2j)}\epsilon^{2j},\quad s=1+\sum_{j=1}^{\infty}s^{(2j)}\epsilon^{2j},
        \label{generalexpansion}
 \end{align}
and substitute them into the Hirota bilinear forms (\ref{HirotaGIandCLL-a})-(\ref{HirotaGIandCLL-d}) and (\ref{HirotaKN-a})-(\ref{HirotaKN-d}). To find $N$-soliton solutions of the system (\ref{cKundu-a}) and (\ref{cKundu-b}), we take
\begin{align}\displaystyle
&g^{(1)}=\sum_{j=1}^N e^{\theta_j},\quad \theta_j=k_jx+\omega_jt+\alpha_j,\, j=1,\ldots, N,\label{generalg_1}\\
&h^{(1)}=\sum_{j=1}^N e^{\eta_j},\quad \eta_j=l_jx+m_jt+\delta_j,\, j=1,\ldots, N,\label{generalh_1}
\end{align}
where $k_j, l_j, \omega_j, m_j, \alpha_j$, and $\delta_j$ are constants.

\subsection{One-soliton solutions}

To obtain one-soliton solution of the system (\ref{cKundu-a}) and (\ref{cKundu-b}) for $\beta=-\frac{1}{4}, 0$, and $\beta=-\frac{1}{2}$, we take $N=1$ in
(\ref{generalg_1}) and (\ref{generalh_1}) that is $g^{(1)}=e^{\theta_1}$ and $h^{(1)}=e^{\eta_1}$, where $\theta_1=k_1x+\omega_1t+\alpha_1$ and $\eta_1=l_1x+m_1t+\delta_1$. Substituting the expansions (\ref{generalexpansion}) into the Hirota bilinear forms (\ref{HirotaGIandCLL-a})-(\ref{HirotaGIandCLL-d}) and (\ref{HirotaKN-a})-(\ref{HirotaKN-d}) with that choice of $g^{(1)}$ and $h^{(1)}$ determines the functions $g$, $h$, $f$ and $s$ as
\begin{equation}
g=\epsilon g^{(1)},\quad h=\epsilon h^{(1)},\quad f=1+\epsilon^2 f^{(2)},\quad s=1+\epsilon^2 s^{(2)}.
\end{equation}
Let us first start with the Hirota bilinear form (\ref{HirotaGIandCLL-a})-(\ref{HirotaGIandCLL-d}) for $\beta=-\frac{1}{4}$ and $\beta=0$ corresponding to CLL, and GI systems, respectively.

\subsubsection{One-soliton solutions of CLL and GI systems}

When we compare the coefficients of the same powers of $\epsilon$ we have
\begin{equation}\displaystyle
\omega_1=\frac{i}{a}k_1^2,\quad m_1=-\frac{i}{a}l_1^2,
\end{equation}
as dispersion relations for both CLL and GI systems. The coefficients of $\epsilon^2$ gives
\begin{equation}\displaystyle
f_2=-\frac{ik_1}{2(k_1+l_1)^2}e^{\theta_1+\eta_1},\quad s_2=\frac{il_1}{2(k_1+l_1)^2}e^{\theta_1+\eta_1}
\end{equation}
for CLL system and
\begin{equation}\displaystyle
f_2=\frac{il_1}{2(k_1+l_1)^2}e^{\theta_1+\eta_1},\quad s_2=-\frac{ik_1}{2(k_1+l_1)^2}e^{\theta_1+\eta_1}
\end{equation}
for GI system. The coefficients of $\epsilon^3$ and $\epsilon^4$ vanish directly. Without loss of generality, take
$\epsilon=1$. Hence one-soliton solution of CLL system (\ref{CLL-a}) and (\ref{CLL-b}) is given by the pair $(q(x,t),r(x,t))$ where
\begin{equation}\label{onesolCLL}
q(x,t)=\frac{e^{k_1x+\omega_1t+\alpha_1}}{1-\frac{ik_1}{2(k_1+l_1)^2}e^{(k_1+l_1)x+(\omega_1+m_1)t+\alpha_1+\delta_1}},\, r(x,t)=\frac{e^{l_1x+m_1t+\delta_1}}{1+\frac{il_1}{2(k_1+l_1)^2}e^{(k_1+l_1)x+(\omega_1+m_1)t+\alpha_1+\delta_1}},
\end{equation}
and one-soliton solution of GI system (\ref{GI-a}) and (\ref{GI-b}) is
\begin{equation}\label{onesolGI}
q(x,t)=\frac{e^{k_1x+\omega_1t+\alpha_1}}{1+\frac{il_1}{2(k_1+l_1)^2}e^{(k_1+l_1)x+(\omega_1+m_1)t+\alpha_1+\delta_1}},\, r(x,t)=\frac{e^{l_1x+m_1t+\delta_1}}{1-\frac{ik_1}{2(k_1+l_1)^2}e^{(k_1+l_1)x+(\omega_1+m_1)t+\alpha_1+\delta_1}},
\end{equation}
with $\omega_1=\frac{i}{a}k_1^2, m_1=-\frac{i}{a}l_1^2$. Here $k_1, l_1, \alpha_1, \delta_1$, and $a$ are arbitrary constants.

\subsubsection{One-soliton solution of KN system}

To find one-soliton solution of KN system (\ref{KN-a}) and (\ref{KN-b}), we insert the expansion (\ref{generalexpansion}) with (\ref{generalg_1})
and (\ref{generalh_1}) for $N=1$ into the Hirota bilinear form of KN system given by (\ref{HirotaKN-a})-(\ref{HirotaKN-d}). Then we make the coefficients of the powers of $\epsilon$ equal to zero. The dispersion relations, the functions $f_2$ and $s_2$ for one-soliton solutions of KN system are same with CLL system (\ref{CLL-a}) and (\ref{CLL-b}). However, the bilinearizing transformation for KN system is different than the one for CLL system. Therefore the forms of the solutions are different. One-soliton solution of KN system (\ref{KN-a}) and (\ref{KN-b}) is given by the pair $(q(x,t),r(x,t))$,
\begin{align}\displaystyle
&q(x,t)=\frac{e^{k_1x+\omega_1t+\alpha_1}\Big[1+\frac{i}{2}\frac{l_1}{(k_1+l_1)^2}e^{(k_1+l_1)x+(\omega_1+m_1)t+\alpha_1+\delta_1}\Big]}
{\Big[1-\frac{i}{2}\frac{k_1}{(k_1+l_1)^2}e^{(k_1+l_1)x+(\omega_1+m_1)t+\alpha_1+\delta_1}\Big]^2},\\
&r(x,t)=\frac{e^{l_1x+m_1t+\delta_1}\Big[1-\frac{i}{2}\frac{k_1}{(k_1+l_1)^2}e^{(k_1+l_1)x+(\omega_1+m_1)t+\alpha_1+\delta_1}\Big]}
{\Big[1+\frac{i}{2}\frac{l_1}{(k_1+l_1)^2}e^{(k_1+l_1)x+(\omega_1+m_1)t+\alpha_1+\delta_1}\Big]^2},
\end{align}
where $\omega_1=i\frac{k_1^2}{a}t$, $m_1=-i\frac{l_1^2}{a}t$. Here $k_1,l_1,\alpha_1,\delta_1,a$ are arbitrary complex numbers.

\subsection{Two-soliton solutions}

To find two-soliton solution of the system (\ref{cKundu-a}) and (\ref{cKundu-b}) for $\beta=-\frac{1}{4}, 0$, and $\beta=-\frac{1}{2}$, we take $N=2$ in
(\ref{generalg_1}) and (\ref{generalh_1}) that is $g^{(1)}=e^{\theta_1}+e^{\theta_2}$ and $h^{(1)}=e^{\eta_1}+e^{\eta_2}$, where $\theta_j=k_jx+\omega_jt+\alpha_j$ and $\eta_j=l_jx+m_jt+\delta_j$ for $j=1, 2$. When we substitute the expansions (\ref{generalexpansion}) into the Hirota bilinear forms (\ref{HirotaGIandCLL-a})-(\ref{HirotaGIandCLL-d}) and (\ref{HirotaKN-a})-(\ref{HirotaKN-d}) with that choice of $g^{(1)}$ and $h^{(1)}$ we obtain the functions $g$, $h$, $f$, and $s$ as
\begin{equation}\label{twosol-func}
g=\epsilon g^{(1)}+\epsilon^3g^{(3)},\quad h=\epsilon h^{(1)}+\epsilon^3h^{(3)},\quad f=1+\epsilon^2 f^{(2)}+\epsilon^4 f^{(4)},\quad s=1+\epsilon^2 s^{(2)}+\epsilon^4 s^{(4)}.
\end{equation}
Now let us first consider two-soliton solutions of CLL and GI systems, and then KN system.

\subsubsection{Two-soliton solutions of CLL and GI systems}

When we make the coefficients of the powers of $\epsilon$ vanish, we get
\begin{equation}\displaystyle
\omega_j=\frac{i}{a}k_j^2,\quad m_j=-\frac{i}{a}l_j^2,\quad j=1, 2
\end{equation}
as the dispersion relations. The coefficients of $\epsilon^2$ give
\begin{equation}\displaystyle
f_2=-\frac{i}{2}\sum_{1\leq m, n\leq 2} \frac{k_m}{(k_m+l_n)^2}e^{\theta_m+\eta_n}, \quad s_2=\frac{i}{2}\sum_{1\leq m, n\leq 2} \frac{l_n}{(k_m+l_n)^2}e^{\theta_m+\eta_n}
\end{equation}
for CLL system and
\begin{equation}\displaystyle
f_2=\frac{i}{2}\sum_{1\leq m, n\leq 2} \frac{l_n}{(k_m+l_n)^2}e^{\theta_m+\eta_n}, \quad s_2=-\frac{i}{2}\sum_{1\leq m, n\leq 2} \frac{k_m}{(k_m+l_n)^2}e^{\theta_m+\eta_n}
\end{equation}
for GI system. The coefficients of $\epsilon^3$ yield
\begin{equation}
g_3=\gamma_1 e^{\theta_1+\theta_2+\eta_1}+\gamma_2 e^{\theta_1+\theta_2+\eta_2},\quad h_3=\beta_1 e^{\theta_1+\eta_1+\eta_2}+\beta_2 e^{\theta_2+\eta_1+\eta_2}
\end{equation}
for both CLL and GI systems, where
\begin{equation}\label{gammabeta}\displaystyle
\gamma_j=\frac{il_j(k_1-k_2)^2}{2(k_1+l_j)^2(k_2+l_j)^2},\quad \beta_j=-\frac{ik_j(l_1-l_2)^2}{2(k_j+l_1)^2(k_j+l_2)^2},\quad j=1, 2.
\end{equation}
We obtain the functions $f_4$ and $s_4$ from the coefficients of $\epsilon^4$ as
\begin{equation}
f_4=M_1e^{\theta_1+\theta_2+\eta_1+\eta_2},\quad g_4=M_2e^{\theta_1+\theta_2+\eta_1+\eta_2}
\end{equation}
for CLL system, and
\begin{equation}
f_4=M_2e^{\theta_1+\theta_2+\eta_1+\eta_2},\quad g_4=M_1e^{\theta_1+\theta_2+\eta_1+\eta_2}
\end{equation}
for GI system, where
\begin{align}\label{M_1M_2}\displaystyle
&M_1=-\frac{k_1k_2(l_1-l_2)^2(k_1-k_2)^2}{4(k_1+l_1)^2(k_1+l_2)^2(k_2+l_1)^2(k_2+l_2)^2},\nonumber\\ &M_2=-\frac{l_1l_2(l_1-l_2)^2(k_1-k_2)^2}{4(k_1+l_1)^2(k_1+l_2)^2(k_2+l_1)^2(k_2+l_2)^2}.
\end{align}
The coefficients of $\epsilon^n$, $n=5, 6, 7, 8$ vanish directly. Take also $\epsilon=1$. Hence two-soliton solution of CLL system (\ref{CLL-a}) and (\ref{CLL-b})
is given by the pair $(q(x,t),r(x,t))$,
\begin{align}\displaystyle
&q(x,t)=\frac{e^{\theta_1}+e^{\theta_2}+\gamma_1e^{\theta_1+\theta_2+\eta_1}+\gamma_2e^{\theta_1+\theta_2+\eta_2}}{1-\frac{i}{2}\sum_{1\leq m,n\leq2}\frac{k_m}{(k_m+l_n)^2}e^{\theta_m+\eta_n}+M_1e^{\theta_1+\theta_2+\eta_1+\eta_2}},\label{twoq(x,t)}\\
&r(x,t)=\frac{e^{\eta_1}+e^{\eta_2}+\beta_1e^{\theta_1+\eta_1+\eta_2}+\beta_2e^{\theta_2+\eta_1+\eta_2}}{1+\frac{i}{2}\sum_{1\leq m,n\leq2}\frac{l_m}{(k_m+l_n)^2}e^{\theta_m+\eta_n}+M_2e^{\theta_1+\theta_2+\eta_1+\eta_2}},\label{twor(x,t)}
\end{align}
where $\theta_j=k_jx+\frac{i}{a}k_j^2t+\alpha_j$ and $\eta_j=l_jx-\frac{i}{a}l_j^2t+\delta_j$ for $j=1, 2$. If we interchange the denominators of the functions $q(x,t)$ and $r(x,t)$ in (\ref{twoq(x,t)}) and (\ref{twor(x,t)}), we get two-soliton solution of GI system (\ref{GI-a}) and (\ref{GI-b}).

\subsubsection{Two-soliton solution of KN system}

Similar to one-soliton solution case, the functions in the expansion (\ref{twosol-func}) for two-soliton solution are same for both CLL and KN systems. When we insert these functions into the bilinearizing transformation (\ref{biltranKN}) for KN system (\ref{KN-a}) and (\ref{KN-b}) we obtain
two-soliton solution of KN system as{\small
\begin{align}
&q(x,t)=\frac{[e^{\theta_1}+e^{\theta_2}+\gamma_1e^{\theta_1+\theta_2+\eta_1}+\gamma_2e^{\theta_1+\theta_2+\eta_2}][1+\frac{i}{2}\sum_{1\leq m,n\leq2}\frac{l_m}{(k_m+l_n)^2}e^{\theta_m+\eta_n}+M_2e^{\theta_1+\theta_2+\eta_1+\eta_2}]}{[1-\frac{i}{2}\sum_{1\leq m,n\leq2}\frac{k_m}{(k_m+l_n)^2}e^{\theta_m+\eta_n}+M_1e^{\theta_1+\theta_2+\eta_1+\eta_2}]^2},\\
&r(x,t)=\frac{[e^{\eta_1}+e^{\eta_2}+\beta_1e^{\theta_1+\eta_1+\eta_2}+\beta_2e^{\theta_2+\eta_1+\eta_2}][1-\frac{i}{2}\sum_{1\leq m,n\leq2}\frac{k_m}{(k_m+l_n)^2}e^{\theta_m+\eta_n}+M_1e^{\theta_1+\theta_2+\eta_1+\eta_2}]}{[1+\frac{i}{2}\sum_{1\leq m,n\leq2}\frac{l_m}{(k_m+l_n)^2}e^{\theta_m+\eta_n}+M_2e^{\theta_1+\theta_2+\eta_1+\eta_2}]^2},
\end{align}}
where $\theta_j=k_jx+\frac{i}{a}k_j^2t+\alpha_j$ and $\eta_j=l_jx-\frac{i}{a}l_j^2t+\delta_j$, $j=1, 2$, $\gamma_j$ and $\beta_j$, $j=1,2$ are given in (\ref{gammabeta}), $M_1$ and $M_2$ are given in (\ref{M_1M_2}).

\section{Reductions}

\noindent A.\, \textbf{Local reductions}

\noindent (i)\, $r(x,t)=\rho \bar{q}(x,t)$, $\rho \in \mathbb{R}$.

By this local reduction, the system (\ref{cKundu-a}) and (\ref{cKundu-b}) reduces to Kundu type equation,
\begin{align}\label{redlocal}
&aq_t(x,t)-iq_{xx}(x,t)+(4\beta+1)\rho q^2(x,t)\bar{q}_x(x,t)+4\beta \rho \bar{q}(x,t)q(x,t)q_x(x,t)\nonumber\\
&-\frac{i}{2}(1+2\beta)(4\beta+1)\rho^2q^3(x,t)\bar{q}^2(x,t)=0,
\end{align}
which is also known as generalized mixed NLS equation. Here for consistency we have $a=\bar{a}$. For $\beta=-\frac{1}{4}$ we have local CLL equation
\begin{equation}\label{redlocalCLL}
aq_t(x,t)-iq_{xx}(x,t)- \rho \bar{q}(x,t)q(x,t)q_x(x,t)=0,
\end{equation}
for $\beta=0$ we get local GI equation
\begin{equation}\label{redlocalGI}
aq_t(x,t)-iq_{xx}(x,t)+\rho q^2(x,t)\bar{q}_x(x,t)-\frac{i}{2}\rho^2q^3(x,t)\bar{q}^2(x,t)=0,
\end{equation}
and for $\beta=-\frac{1}{2}$ we obtain local KN equation
\begin{equation}\label{redlocalKN}
aq_t(x,t)-iq_{xx}(x,t)-\rho q^2(x,t)\bar{q}_x(x,t)-2\rho \bar{q}(x,t)q(x,t)q_x(x,t)=0.
\end{equation}

\noindent B.\, \textbf{Shifted nonlocal reductions}

\noindent (i)\, $r(x,t)=\rho q(-x+x_0,-t+t_0)$, $\rho, x_0, t_0 \in \mathbb{R}$.

Under this reduction, the system (\ref{cKundu-a}) and (\ref{cKundu-b}) consistently reduces to the real shifted nonlocal space-time reversal equation,
\begin{align}\label{redrealSTrev}
aq_t(x,t)-iq_{xx}(x,t)+(4\beta&+1)\rho q^2(x,t)q_x(-x+x_0,-t+t_0)\nonumber\\
&+4\beta\rho q(x,t)q_x(x,t)q(-x+x_0,-t+t_0)\nonumber\\
&\hspace{0.5cm}-\frac{i}{2}(1+2\beta)(4\beta+1)\rho^2q^3(x,t)q^2(-x+x_0,-t+t_0)=0,
\end{align}
for any $\beta \in \mathbb{R}$. For $\beta=-\frac{1}{4}$ we have real shifted nonlocal space-time reversal CLL equation
\begin{equation}\label{redrealSTrevCLL}
aq_t(x,t)-iq_{xx}(x,t)- \rho q(x,t)q_x(x,t)q(-x+x_0,-t+t_0)=0.
\end{equation}
When $x_0=t_0=0$, we get real standard nonlocal space-time reversal CLL equation \cite{YangYang}. For $\beta=0$ we obtain real shifted nonlocal space-time reversal GI equation
\begin{equation}\label{redrealSTrevGI}
aq_t(x,t)-iq_{xx}(x,t)+\rho q^2(x,t)q_x(-x+x_0,-t+t_0)-\frac{i}{2}\rho^2q^3(x,t)q^2(-x+x_0,-t+t_0)=0.
\end{equation}
In \cite{LZYY}, soliton solutions for the above equation when $x_0=t_0=0$ that is for real standard nonlocal
space-time reversal GI equation have been obtained by Darboux transformation.

For $\beta=-\frac{1}{2}$ we obtain the real shifted nonlocal space-time reversal KN equation
\begin{equation}\label{redrealSTrevKN}
aq_t(x,t)-iq_{xx}(x,t)-\rho q^2(x,t)q_x(-x+x_0,-t+t_0)-2\rho q(x,t)q(-x+x_0,-t+t_0)q_x(x,t)=0.
\end{equation}
In \cite{MuGu}, Mukherjee and Guha derived the nonholonomic deformation of the real standard ($x_0=t_0=0$) nonlocal space-time reversal coupled
KN system. Ablowitz and Musslimani gave Lax pairs and the first two conserved quantities of the standard unshifted form of the above equation in \cite{AbMu3}. Valchev presented multi-component version of the real standard nonlocal space-time reversal KN system as an example for the generalization of Mikhailov's reduction group method \cite{Valchev}.\\

\noindent (ii)\, $r(x,t)=\rho \bar{q}(x,-t+t_0)$, $\rho, t_0 \in \mathbb{R}$.

When we apply this reduction to the system (\ref{cKundu-a}) and (\ref{cKundu-b}), for consistency we have $a=-\bar{a}$ and the system reduces to the complex shifted nonlocal time reversal equation,
\begin{align}\label{redcompTrev}
&aq_t(x,t)-iq_{xx}(x,t)+(4\beta+1)\rho q^2(x,t)\bar{q}_x(x,-t+t_0)+4\beta\rho q(x,t)q_x(x,t)\bar{q}(x,-t+t_0)\nonumber\\
&-\frac{i}{2}(1+2\beta)(4\beta+1)\rho^2q^3(x,t)\bar{q}^2(x,-t+t_0)=0,
\end{align}
for any $\beta \in \mathbb{R}$. For $\beta=-\frac{1}{4}$ we have complex shifted nonlocal time reversal CLL equation
\begin{equation}\label{redcompTrevCLL}
aq_t(x,t)-iq_{xx}(x,t)-\rho q(x,t)q_x(x,t)\bar{q}(x,-t+t_0)=0,
\end{equation}
for $\beta=0$ we get complex shifted nonlocal time reversal GI equation
\begin{equation}\label{redcompTrevGI}
aq_t(x,t)-iq_{xx}(x,t)+\rho q^2(x,t)\bar{q}_x(x,-t+t_0)-\frac{i}{2}\rho^2q^3(x,t)\bar{q}^2(x,-t+t_0)=0,
\end{equation}
and for $\beta=-\frac{1}{2}$ we obtain complex shifted nonlocal time reversal KN equation
\begin{equation}\label{redcompTrevKN}
aq_t(x,t)-iq_{xx}(x,t)-\rho q^2(x,t)\bar{q}_x(x,-t+t_0)-2\rho q(x,t)q_x(x,t)\bar{q}(x,-t+t_0)=0.
\end{equation}
In \cite{Zhou}, Zhou obtained soliton solutions for the above equation when $t_0=0$ that is for complex standard unshifted nonlocal time reversal KN equation by Darboux transformation.\\

\noindent (iii)\, $r(x,t)=\rho q(-x+x_0,t)$, $\rho, x_0 \in \mathbb{R}$.

Applying this reduction to the system (\ref{cKundu-a}) and (\ref{cKundu-b}) gives $a=0$ and the system consistently reduces to the real stationary shifted nonlocal space reversal equation,
\begin{align}\label{redrealSrev}
&iq_{xx}(x,t)-(4\beta+1)\rho q^2(x,t)q_x(-x+x_0,t)-4\beta\rho q(x,t)q_x(x,t)q(-x+x_0,t)\nonumber\\
&+\frac{i}{2}(1+2\beta)(4\beta+1)\rho^2q^3(x,t)q^2(-x+x_0,t)=0,
\end{align}
for any $\beta \in \mathbb{R}$. In this paper we will not deal with the solutions of this stationary equation.

\section{Soliton solutions of the local and shifted nonlocal reduced equations}

By using the reduction formulas with one- and two-soliton solutions of (\ref{cKundu-a}) and (\ref{cKundu-b}) for particular values of $\beta$ and following Type 1 or Type 2 approaches \cite{GurPek1}, \cite{GurPek3}, \cite{GurPek2} we can also obtain one- and two-soliton solutions of the local and shifted nonlocal reduced Kundu type equations.

\subsection{One-soliton solutions of the local reduced equations}

In this section we will first consider the local reduced CLL and GI equations, then the
local reduced KN equation.

\subsubsection{One-soliton solutions of the local reduced CLL and GI equations}

\noindent A.(i)\, $r(x,t)=\rho \bar{q}(x,t), a=\bar{a}$, $\rho\in \mathbb{R}$.

Using Type 1 approach based on equating numerators and denominators
separately in the reduction formula with one-soliton solutions of CLL and GI systems, we get the following constraints:
\begin{equation}
l_1=\bar{k}_1,\quad e^{\delta_1}=\rho e^{\bar{\alpha}_1}.
\end{equation}
Therefore one-soliton solution of the local reduced CLL equation (\ref{redlocalCLL}) is
\begin{equation}\label{local1SSCLL}\displaystyle
q(x,t)=\frac{e^{k_1x+\omega_1t+\alpha_1}}{1-\frac{i\rho k_1}{2(k_1+\bar{k}_1)^2}e^{(k_1+\bar{k}_1)x+(\omega_1+\bar{\omega}_1)t+\alpha_1+\bar{\alpha}_1}},
\end{equation}
 and one-soliton solution of the local reduced GI equation (\ref{redlocalGI}) is
 \begin{equation}\label{local1SSGI}\displaystyle
q(x,t)=\frac{e^{k_1x+\omega_1t+\alpha_1}}{1+\frac{i\rho \bar{k}_1}{2(k_1+\bar{k}_1)^2}e^{(k_1+\bar{k}_1)x+(\omega_1+\bar{\omega}_1)t+\alpha_1+\bar{\alpha}_1}},
\end{equation}
where $\omega_1=\frac{i}{a}k_1^2$.

Note that if $k_1$ is a pure imaginary number then the solutions (\ref{local1SSCLL}) and (\ref{local1SSGI}) become both trivial solutions $q(x,t)=0$. Let now $k_1=\xi_1+i\mu_1$, $\omega_1=\xi_2+i\mu_2$, and $e^{\alpha_1}=\xi_3+i\mu_3$ for $\xi_j, \mu_j \in \mathbb{R}$, $j=1, 2, 3$ in the solutions (\ref{local1SSCLL}) and (\ref{local1SSGI}). Here $\xi_2=-\frac{2}{a}\xi_1\mu_1$ and $\mu_2=\frac{\xi_1^2-\mu_1^2}{a}$. We get the same real-valued solution $|q(x,t)|^2$
for both local reduced CLL and GI equations as
\begin{equation}\label{local1SSCLLGIKN}\displaystyle
|q(x,t)|^2=\frac{4\sigma_1\xi_1^2}{\rho\sqrt{\xi_1^2+\mu_1^2}[\cosh(2\xi_1x+2\xi_2t+\delta)+\frac{\mu_1\sigma_1}{\sqrt{\xi_1^2+\mu_1^2}}  ]},
\end{equation}
where $\delta=\ln\Big|\frac{\rho \sqrt{\xi_1^2+\mu_1^2}(\xi_3^2+\mu_3^2)}{8\xi_1^2}\Big|$, $\sigma_1=\pm 1$. The solution (\ref{local1SSCLLGIKN})
is nonsingular for $\frac{\mu_1\sigma_1}{\sqrt{\xi_1^2+\mu_1^2}}>-1$.

\subsubsection{One-soliton solution of the local reduced KN equation}

\noindent A.(i)\, $r(x,t)=\rho \bar{q}(x,t), a=\bar{a}$, $\rho\in \mathbb{R}$.

Using Type 1 with the reduction formula and one-soliton solution of KN system gives the constraints
\begin{equation}
l_1=\bar{k}_1,\quad e^{\delta_1}=\rho e^{\bar{\alpha}_1}.
\end{equation}
Hence one-soliton solution of the local reduced KN equation (\ref{redlocalKN}) is
\begin{equation}\label{local1SSKN}\displaystyle
q(x,t)=\frac{e^{k_1x+\omega_1t+\alpha_1}[1+\frac{i\rho \bar{k}_1}{2(k_1+\bar{k}_1)^2}e^{(k_1+\bar{k}_1)x+(\omega_1+\bar{\omega}_1)t+\alpha_1+\bar{\alpha}_1}   ]}
{[1-\frac{i\rho k_1}{2(k_1+\bar{k}_1)^2}e^{(k_1+\bar{k}_1)x+(\omega_1+\bar{\omega}_1)t+\alpha_1+\bar{\alpha}_1} ]^2},
\end{equation}
where $\omega_1=\frac{ik_1^2}{a}$ and $k_1, \alpha_1$ are arbitrary constants. Here if $k_1$ is a pure imaginary number then the solution (\ref{local1SSKN}) become a trivial solution $q(x,t)=0$. Additionally, the solution (\ref{local1SSKN}) yields the same solution $|q(x,t)|^2$ given by (\ref{local1SSCLLGIKN}).

\noindent \textbf{Example 1.} Let us take the parameters of one-soliton solution $|q(x,t)|^2$ of the local reduced CLL, GI, and KN equations as
$k_1=1+\frac{i}{2}, a=\rho=e^{\alpha_1}=1$. Then we get the one-soliton solution as
\begin{equation}\displaystyle
|q(x,t)|^2=\frac{8}{\sqrt{5}[\cosh(2x-2t+\delta)+\frac{1}{\sqrt{5}}]},
\end{equation}
where $\delta=\ln\Big|\frac{16}{\sqrt{5}}\Big|$. The graph of the above nonsingular solution is given in
Figure 1.
\begin{center}
\begin{figure}[h]
\centering
\begin{minipage}[t]{1\linewidth}
\centering
\includegraphics[angle=0,scale=.33]{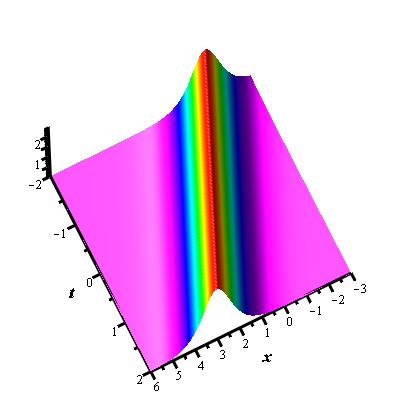}
\caption{One-soliton solution with the parameters $k_1=1+\frac{i}{2}, a=\rho=e^{\alpha_1}=1$ for the local reduced CLL, GI, and KN equations.}
\end{minipage}%
\end{figure}
\end{center}
\squeezeup

\subsection{One-soliton solutions of the shifted nonlocal reduced equations}

Here at first we will consider the shifted nonlocal reduced CLL and GI equations and find one-soliton solutions of these
reduced equations. Secondly, we will find one-soliton solutions of the shifted nonlocal
reduced KN equations.

\subsubsection{One-soliton solutions of the shifted nonlocal reduced CLL and GI equations}

\noindent B.(i)\, $r(x,t)=\rho q(-x+x_0,-t+t_0)$, $\rho, x_0, t_0 \in \mathbb{R}$.

Here using Type 1 gives $l_1=-k_1$ yielding trivial solutions for both shifted nonlocal space-time reversal reduced CLL and GI equations.
Therefore we apply Type 2 approach which is based on the
cross multiplication of the reduction formula and get the following constraints:
\begin{equation}\displaystyle
e^{\alpha_1}=\sigma_1\frac{\sqrt{2}(k_1+l_1)}{\sqrt{i\rho l_1}}e^{\frac{-k_1x_0-\omega_1t_0}{2}},\, e^{\delta_1}=\sigma_2\frac{\sqrt{2}\sqrt{\rho}(k_1+l_1)}{\sqrt{-i k_1}}e^{\frac{-l_1x_0-m_1t_0}{2}},\, \sigma_j=\pm 1,\, j=1, 2
\end{equation}
for the shifted nonlocal space-time reversal reduced CLL equation (\ref{redrealSTrevCLL}). Then one-soliton solution of this equation becomes
\begin{equation}\label{1SSnonlocaliCLL}\displaystyle
q(x,t)=\sigma_1\sqrt{\frac{-2i}{\rho l_1}}\frac{(k_1+l_1)e^{k_1x+\omega_1t-\frac{(k_1x_0+\omega_1t_0)}{2}}}{[1-i\frac{\sqrt{k_1}}{\sqrt{l_1}}\sigma_1\sigma_2e^{(k_1+l_1)x+(\omega_1+m_1)t
-\frac{((k_1+l_1)x_0+(\omega_1+m_1)t_0)}{2}} ]},
\end{equation}
where $\omega_1=\frac{i}{a}k_1^2, m_1=-\frac{i}{a}l_1^2$ and $\sigma_j=\pm 1$, $j=1, 2$. Note that if $u(x,t)$ is the one-soliton solution of  the standard unshifted nonlocal space-time reversal reduced CLL equation obtained by Type 2 then we have the relation $q(x,t)=u(x-\frac{x_0}{2},t-\frac{t_0}{2})$ satisfied.

Let $k_1, l_1\in \mathbb{R}$ and $a=i\xi_1$, $\xi_1\in \mathbb{R}$ in (\ref{1SSnonlocaliCLL}). In this case,
we have $\omega_1=\frac{k_1^2}{\xi_1}$ and $m_1=-\frac{l_1^2}{\xi_1}$. Then we get the following real-valued solution for the shifted nonlocal space-time reversal reduced CLL equation (\ref{redrealSTrevCLL}) as
\begin{equation}\label{mod1SSnonlocaliCLL}
|q(x,t)|^2=\frac{2(k_1+l_1)^2e^{2k_1x+2\omega_1t-(k_1x_0+\omega_1t_0)}}{\rho l_1[1+\frac{k_1}{l_1}e^{2(k_1+l_1)x+(\omega_1+m_1)t-((k_1+l_1)x_0+(\omega_1+m_1)t_0)}  ]}.
\end{equation}
The above solution is nonsingular if $\frac{k_1}{l_1}\geq 0$.

\noindent \textbf{Example 2.} Choose the parameters of the solution (\ref{mod1SSnonlocaliCLL}) as
$k_1=1, l_1=\frac{1}{2}, a=2i, \sigma_1=\sigma_2=\rho=x_0=1, t_0=-1$. So one-soliton solution of the equation (\ref{redrealSTrevCLL}) becomes
\begin{equation}\displaystyle
|q(x,t)|^2=\frac{9}{2\sqrt{2}}e^{\frac{1}{2}x+\frac{5}{8}t+\frac{1}{16}}\mathrm{sech}(\frac{3}{2}x+\frac{3}{8}t-\frac{9}{16}+\frac{1}{2}\ln 2).
\end{equation}
This is an asymptotically decaying nonsingular solution. The graph of the above solution is given in
Figure 2.
\begin{center}
\begin{figure}[h]
\centering
\begin{minipage}[t]{1\linewidth}
\centering
\includegraphics[angle=0,scale=.33]{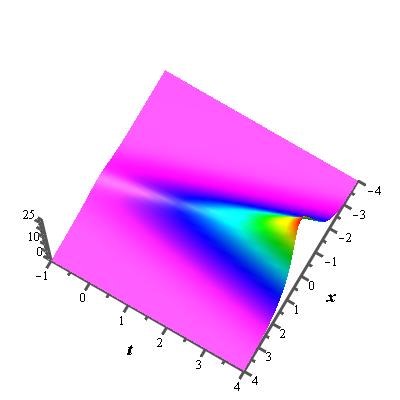}
\caption{Asymptotically decaying solution with the parameters $k_1=1, l_1=\frac{1}{2}, a=2i, \sigma_1=\sigma_2=\rho=x_0=1, t_0=-1$ for the shifted nonlocal space-time reversal reduced CLL equation.}
\end{minipage}%
\end{figure}
\end{center}
\squeezeup

For the shifted nonlocal space-time reversal reduced GI equation (\ref{redrealSTrevGI}), Type 2 gives
\begin{equation}\displaystyle
e^{\alpha_1}=\sigma_1\frac{\sqrt{2}(k_1+l_1)}{\sqrt{-i\rho k_1}}e^{\frac{-k_1x_0-\omega_1t_0}{2}},\, e^{\delta_1}=\sigma_2\frac{\sqrt{2}\sqrt{\rho}(k_1+l_1)}{\sqrt{i l_1}}e^{\frac{-l_1x_0-m_1t_0}{2}},\, \sigma_j=\pm 1,\, j=1, 2.
\end{equation}
Hence one-soliton solution of the equation (\ref{redrealSTrevGI}) is
\begin{equation}\label{1SSnonlocaliGI}\displaystyle
q(x,t)=\sigma_1\sqrt{\frac{2i}{\rho k_1}}\frac{(k_1+l_1)e^{k_1x+\omega_1t-\frac{(k_1x_0+\omega_1t_0)}{2}}}{[1+i\frac{\sqrt{l_1}}{\sqrt{k_1}}\sigma_1\sigma_2e^{(k_1+l_1)x+(\omega_1+m_1)t
-\frac{((k_1+l_1)x_0+(\omega_1+m_1)t_0)}{2}} ]},
\end{equation}
where $\omega_1=\frac{i}{a}k_1^2, m_1=-\frac{i}{a}l_1^2$ and $\sigma_j=\pm 1$, $j=1, 2$.  Note that if $u(x,t)$ is the one-soliton solution of  the standard unshifted nonlocal space-time reversal reduced GI equation obtained by Type 2 then we have the relation $q(x,t)=u(x-\frac{x_0}{2},t-\frac{t_0}{2})$ satisfied.

Similar to the shifted nonlocal space-time reversal reduced CLL equation, if we take $k_1, l_1\in \mathbb{R}$ and $a=i\xi_1$, $\xi_1\in \mathbb{R}$
in (\ref{1SSnonlocaliGI}), we get
the real-valued solution for the shifted nonlocal space-time reversal reduced GI equation (\ref{redrealSTrevGI}) as
\begin{equation}\label{mod1SSnonlocaliGI}\displaystyle
|q(x,t)|^2=\frac{2(k_1+l_1)^2e^{2k_1x+2\omega_1t-(k_1x_0+\omega_1t_0)}}{\rho k_1[1+\frac{l_1}{k_1}e^{2(k_1+l_1)x+(\omega_1+m_1)t-((k_1+l_1)x_0+(\omega_1+m_1)t_0)}  ]}.
\end{equation}
The above solution is nonsingular if $\frac{l_1}{k_1}\geq 0$. If $k_1=l_1$, the solutions (\ref{mod1SSnonlocaliCLL}) and (\ref{mod1SSnonlocaliGI}) become same. If we take $x_0=t_0=0$, the solution (\ref{mod1SSnonlocaliGI}) turns to the same solution which was obtained by Darboux transformation represented in \cite{LZYY} given for the real standard unshifted nonlocal  space-time reversal GI equation.

\noindent \textbf{Example 3.} Take the parameters of the solution (\ref{1SSnonlocaliGI}) as
$k_1=i, l_1=3i, a=3, \sigma_1=\sigma_2=\rho=x_0=1, t_0=-2$. Hence one-soliton solution of the equation (\ref{redrealSTrevGI}) becomes
\begin{equation}\displaystyle
|q(x,t)|^2=\frac{16}{2-\sqrt{3}\sin(4x+\frac{8}{3}t+\frac{2}{3} )}.
\end{equation}
This is a periodic wave solution. The graph of the solution is given in
Figure 3.
\begin{center}
\begin{figure}[h]
\centering
\begin{minipage}[t]{1\linewidth}
\centering
\includegraphics[angle=0,scale=.33]{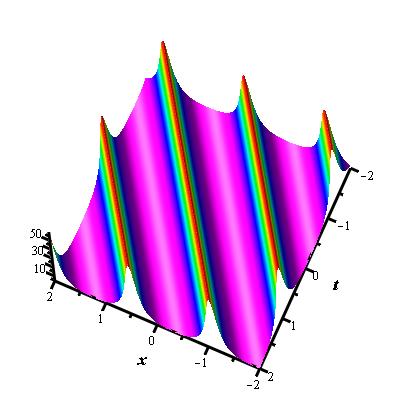}
\caption{Periodic wave solution with the parameters $k_1=i, l_1=3i, a=3, \sigma_1=\sigma_2=\rho=x_0=1, t_0=-2$ for the shifted nonlocal space-time reversal reduced GI equation.}
\end{minipage}%
\end{figure}
\end{center}
\squeezeup

\noindent B.(ii)\, $r(x,t)=\rho \bar{q}(x,-t+t_0)$, $\rho, t_0 \in \mathbb{R}$, $a=-\bar{a}$.

When we use Type 1 with the reduction formula $r(x,t)=\rho \bar{q}(x,-t+t_0)$ and one-soliton solutions (\ref{onesolCLL}) and (\ref{onesolGI}),
we get the constraints,
\begin{equation}
l_1=\bar{k}_1,\quad e^{\delta_1}=\rho e^{\bar{\omega}_1t_0+\bar{\alpha}_1},
\end{equation}
giving $m_1=-\bar{\omega}_1$. Hence one-soliton solution of the complex shifted nonlocal time reversal CLL equation (\ref{redcompTrevCLL}) is
\begin{equation}\label{1SSnonlocaliiCLL}\displaystyle
q(x,t)=\frac{e^{k_1x+\omega_1t+\alpha_1}}{1-\frac{i\rho k_1}{2(k_1+\bar{k}_1)^2}e^{(k_1+\bar{k}_1)x+(\omega_1-\bar{\omega}_1)t}e^{\bar{\omega}_1t_0+\alpha_1+\bar{\alpha}_1}},
\end{equation}
and one-soliton solution of the complex shifted nonlocal time reversal GI equation (\ref{redcompTrevGI}) is
\begin{equation}\label{1SSnonlocaliiGI}\displaystyle
q(x,t)=\frac{e^{k_1x+\omega_1t+\alpha_1}}{1+\frac{i\rho \bar{k}_1}{2(k_1+\bar{k}_1)^2}e^{(k_1+\bar{k}_1)x+(\omega_1-\bar{\omega}_1)t}e^{\bar{\omega}_1t_0+\alpha_1+\bar{\alpha}_1}},
\end{equation}
where $\omega_1=\frac{i}{a}k_1^2, m_1=-\frac{i}{a}l_1^2$. If we compare
the one-soliton solution of the complex standard unshifted nonlocal time reversal CLL and GI equations, say $u(x,t)$, with the above solution obtained by Type 1, we get
$q(x,t)\neq u(x, t+\kappa(t_0))$ for any nonzero function $\kappa$.

Note that if $k_1$ is a pure imaginary number, the solutions (\ref{1SSnonlocaliiCLL})
and (\ref{1SSnonlocaliiGI}) become trivial. Let us take $k_1=\xi_1+i\mu_1$, $\omega_1=\xi_2+i\mu_2$, $e^{\alpha_1}=\xi_3+i\mu_3$, and
$a=i\mu_4$, $\xi_j\in \mathbb{R}$ for $j=1, 2, 3$ and $\mu_j \in \mathbb{R}$ for $j=1, 2, 3, 4$. Here $\xi_2=\frac{1}{\mu_4}(\xi_1^2-\mu_1^2)$, $\mu_2=\frac{2\xi_1\mu_1}{\mu_4}$. Then from both solutions (\ref{1SSnonlocaliiCLL}) and (\ref{1SSnonlocaliiGI}) we get the same
real-valued solution
\begin{equation}\label{mod1SSnonlocalgen}\displaystyle
|q(x,t)|^2=\frac{W_1(x,t)}{W_2(x,t)},
\end{equation}
where
\begin{align}
W_1=&e^{2\xi_1x+2\xi_2t}(\xi_3^2+\mu_3^2)\\
W_2=&1+\frac{\rho(\xi_3^2+\mu_3^2)}{4\xi_1^2}e^{2\xi_1x+\xi_2t_0}[\mu_1\cos(2\mu_2t-\mu_2t_0)+\xi_1\sin(2\mu_2t-\mu_2t_0)]\nonumber\\
&+\frac{\rho^2(\xi_1^2+\mu_1^2)}{64\xi_1^4}(\xi_3^2+\mu_3^2)e^{4\xi_1x+2\xi_2t_0}.
\end{align}
Let us consider a particular case. Take $\xi_1=B\sin\omega_0$ and $\mu_1=B\cos\omega_0\neq 0$ yielding $B^2=\xi_1^2+\mu_1^2$. In this case the above solution becomes
\begin{equation}\displaystyle
|q(x,t)|^2=\frac{2\sigma_1\xi_1^2e^{2\xi_2t_0-\xi_2t_0}}{\rho B[\cosh(2\xi_1x+\xi_2t_0+\delta)+\sigma_1\mu_1\cos(2\mu_2t-\mu_2t_0-\omega_0)]},
\end{equation}
where $\delta=\ln \Big|\frac{\rho\sqrt{\xi_1^2+\mu_1^2}(\xi_3^2+\mu_3^2)}{8\xi_1^2}\Big|$, $\omega_0=\arccos \Big(\frac{\mu_1}{\sqrt{\xi_1^2+\mu_1^2}}\Big)$. The above solution is singular.

If $\mu_1=0$, then $\mu_2=0$ and the real-valued solution (\ref{mod1SSnonlocalgen}) of both complex shifted nonlocal time reversal CLL and GI equations becomes
\begin{equation}\label{mod1SSnonlocalpar}\displaystyle
|q(x,t)|^2=\frac{4\sigma_1\xi_1e^{2\xi_2t-\xi_2t_0}}{\rho}\mathrm{sech}(2\xi_1x+\xi_2t_0+\delta),
\end{equation}
where $\delta=\ln\Big|\frac{\rho(\xi_3^2+\mu_3^2)}{8\xi_1}\Big|$, $\sigma_1=\pm 1$. This solution is nonsingular for any $(x,t)\in \mathbb{R}^2$.

\subsubsection{One-soliton solutions of the shifted nonlocal reduced KN equations}

In this part we will obtain one-soliton solutions of the shifted nonlocal reduced KN equations
by using Type 1 and Type 2 approaches on the reduction formulas with one-soliton solution of the KN equation.

\noindent B.(i)\, $r(x,t)=\rho q(-x+x_0,-t+t_0)$, $\rho, x_0, t_0 \in \mathbb{R}$.

Here using Type 1 gives trivial solution therefore we apply Type 2 approach. We get the following constraints:
\begin{equation}\displaystyle
e^{\alpha_1}=\frac{\sigma_1\sqrt{2ik_1}}{\sqrt{\rho}l_1}(k_1+l_1)e^{\frac{-k_1x_0-\omega_1t_0}{2}},\,\,
e^{\delta_1}=\frac{\sigma_2 \sqrt{-2i\rho l_1}}{k_1}(k_1+l_1)e^{\frac{-l_1x_0-m_1t_0}{2}},
\end{equation}
where $\omega_1=\frac{i}{a}k_1^2, m_1=-\frac{i}{a}l_1^2$, $\sigma_j=\pm 1$, $j=1, 2$. Hence one-soliton solution of the equation (\ref{redrealSTrevKN}) is
\begin{equation}\label{1SSnonlocaliKN}\displaystyle
q(x,t)=\frac{\sigma_1\sqrt{2ik_1}(k_1+l_1)e^{k_1x+\omega_1t-\frac{(k_1x_0+\omega_1t_0)}{2}}
[1+\frac{i\sigma_1\sigma_2\sqrt{l_1}}{\sqrt{k_1}}e^{(k_1+l_1)x+(\omega_1+m_1)t-\frac{((k_1+l_1)x_0+(\omega_1+m_1)t_0)}{2}}  ]}
{\sqrt{\rho}l_1[1-\frac{i\sigma_1\sigma_2\sqrt{k_1}}{\sqrt{l_1}}e^{(k_1+l_1)x+(\omega_1+m_1)t-\frac{((k_1+l_1)x_0+(\omega_1+m_1)t_0)}{2}}  ]^2},
\end{equation}
where $\omega_1=\frac{i}{a}k_1^2, m_1=-\frac{i}{a}l_1^2$ and $\sigma_j=\pm 1$, $j=1, 2$. Note that if $u(x,t)$ is the one-soliton solution of  the standard unshifted nonlocal space-time reversal reduced KN equation obtained by Type 2 then we have the relation $q(x,t)=u(x-\frac{x_0}{2},t-\frac{t_0}{2})$ satisfied.

Let $k_1, l_1\in \mathbb{R}$ and $a=i\xi_1$, $\xi_1\in \mathbb{R}$ yielding $\omega_1=\frac{k_1^2}{\xi_1}$ and $m_1=-\frac{l_1^2}{\xi_1}$. From the solution (\ref{1SSnonlocaliKN}) we obtain the following real-valued solution for the shifted nonlocal space-time reversal reduced KN equation (\ref{redrealSTrevKN}) as
\begin{equation}\label{mod1SSnonlocaliKN}\displaystyle
|q(x,t)|^2=\frac{2k_1(k_1+l_1)^2e^{2k_1x+2\omega_1t-(k_1x_0+\omega_1t_0)[1+\frac{l_1}{k_1}e^{2(k_1+l_1)x+2(\omega_1+m_1)t-((k_1+l_1)x_0+(\omega_1+m_1)t_0)}]}}
{\rho l_1^2[1+\frac{k_1}{l_1}e^{2(k_1+l_1)x+2(\omega_1+m_1)t-((k_1+l_1)x_0+(\omega_1+m_1)t_0)}]^2}.
\end{equation}
This solution is nonsingular if $\frac{k_1}{l_1}\geq 0$. Note that if $k_1=l_1$ then all the solutions (\ref{mod1SSnonlocaliCLL}), (\ref{mod1SSnonlocaliGI}), and (\ref{mod1SSnonlocaliKN}) become same.

\noindent \textbf{Example 4.} Choose the parameters of the solution (\ref{1SSnonlocaliKN}) as
$k_1=i, l_1=3i, a=3, \sigma_1=\sigma_2=\rho=x_0=1, t_0=-2$. Hence one-soliton solution of the equation (\ref{redrealSTrevKN}) becomes $|q(x,t)|^2=\frac{W_1(x,t)}{W_2(x,t)}$,
where
\begin{align}
W_1(x,t)=&640+224\sqrt{3}\sin(4x+\frac{8}{3}t+\frac{2}{3})+384\cos(8x+\frac{16}{3}t+\frac{4}{3})+96\sqrt{3}\sin(12x+8t+2),\nonumber\\
W_2(x,t)=&443+400\sqrt{3}\sin(4x+\frac{8}{3}t+\frac{2}{3})-324\cos(8x+\frac{16}{3}t+\frac{4}{3})-48\sqrt{3}\sin(12x+8t+2)\nonumber\\
&+9\cos(16x+\frac{32}{3}t+\frac{8}{3}).
\end{align}
This is a periodic wave solution. The graph of this nonsingular solution is given in
Figure 4.
\begin{center}
\begin{figure}[h]
\centering
\begin{minipage}[t]{1\linewidth}
\centering
\includegraphics[angle=0,scale=.33]{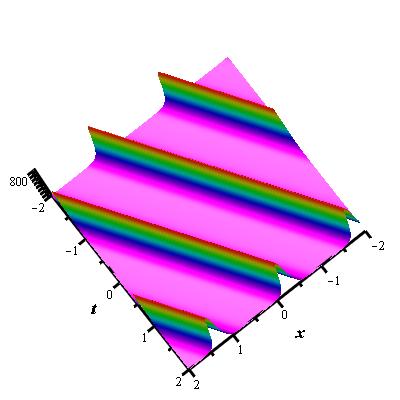}
\caption{Periodic wave solution with the parameters $k_1=i, l_1=3i, a=3, \sigma_1=\sigma_2=\rho=x_0=1, t_0=-2$ for the shifted nonlocal space-time reversal reduced KN equation.}
\end{minipage}%
\end{figure}
\end{center}
\squeezeup

\noindent B.(ii)\, $r(x,t)=\rho \bar{q}(x,-t+t_0)$, $\rho, t_0 \in \mathbb{R}$, $a=-\bar{a}$.

When we apply Type 1 to the reduction formula with one-soliton solution of KN system we get the constraints
\begin{equation}
l_1=\bar{k}_1,\quad e^{\delta_1}=\rho e^{\bar{\omega}_1t_0+\bar{\alpha}_1}.
\end{equation}
Hence one-soliton solution of the complex shifted nonlocal time reversal KN equation (\ref{redcompTrevKN}) is
\begin{equation}\label{1SSnonlocaliiKN}\displaystyle
q(x,t)=\frac{e^{k_1x+\omega_1t+\alpha_1}[1+\frac{i\rho\bar{k}_1}{2(k_1+\bar{k}_1)^2}e^{(k_1+\bar{k}_1)x+(\omega_1-\bar{\omega}_1)t+\alpha_1+\bar{\alpha}_1+\bar{\omega}_1t_0}  ]}{[1-\frac{i\rho k_1}{2(k_1+\bar{k}_1)^2}e^{(k_1+\bar{k}_1)x+(\omega_1-\bar{\omega}_1)t+\alpha_1+\bar{\alpha}_1+\bar{\omega}_1t_0}  ]^2},
\end{equation}
where $\omega_1=\frac{i}{a}k_1^2$. If we compare
the one-soliton solution of the complex standard unshifted nonlocal time reversal KN equation, say $u(x,t)$, with the above solution obtained by Type 1, we get
$q(x,t)\neq u(x, t+\kappa(t_0))$ for any nonzero function $\kappa$.

Similar to the local reduction case, if $k_1$ is a pure imaginary number then all the solutions (\ref{1SSnonlocaliiCLL}), (\ref{1SSnonlocaliiGI}), and (\ref{1SSnonlocaliiKN}) become trivial solutions $q(x,t)=0$. In addition to that we obtain the same real-valued solutions $|q(x,t)|^2$ given by (\ref{mod1SSnonlocalgen}) and (\ref{mod1SSnonlocalpar}) for the complex shifted nonlocal time reversal KN equation (\ref{redcompTrevKN}).

\noindent \textbf{Example 5.} Choose the parameters of the solutions (\ref{1SSnonlocaliiCLL}), (\ref{1SSnonlocaliiGI}), and (\ref{1SSnonlocaliiKN}) as
$k_1=\rho=1, e^{\alpha_1}=-1, a=2i, t_0=2$. Then one-soliton solutions $|q(x,t)|^2$ of the complex shifted nonlocal time reversal CLL, GI, and KN equations become same 
\begin{equation}\displaystyle
|q(x,t)|^2=\frac{4096e^{2x+t}+64e^{6x+t+2}}{4096+128e^{4x+2}+e^{8x+4}}.
\end{equation}
This is an asymptotically decaying solution. The graph of the above nonsingular solution is given in
Figure 5.
\begin{center}
\begin{figure}[h]
\centering
\begin{minipage}[t]{1\linewidth}
\centering
\includegraphics[angle=0,scale=.33]{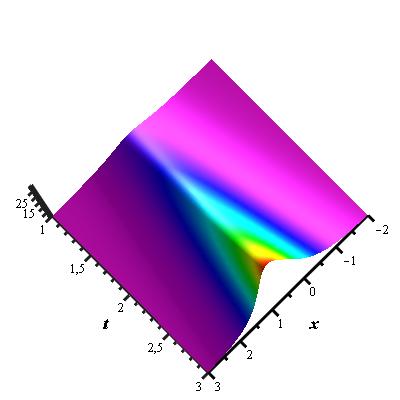}
\caption{Asymptotically decaying solution with the parameters $k_1=\rho=1, e^{\alpha_1}=-1, a=2i, t_0=2$ for the complex shifted nonlocal time reversal reduced CLL, GI, and KN equations.}
\end{minipage}%
\end{figure}
\end{center}
\squeezeup

\subsection{Two-soliton solutions of the local reduced equations}

Here we will obtain two-soliton solutions of the local reduced CLL, GI, and KN equations by following Type 1 and Type 2 approaches.

\subsubsection{Two-soliton solutions of the local reduced CLL and GI equations}

\noindent A.(i)\, $r(x,t)=\rho \bar{q}(x,t), a=\bar{a}$, $\rho\in \mathbb{R}$.

Applying Type 1 to the reduction formula with two-soliton solutions of CLL and GI systems, we obtain the following constraints
on the parameters of two-soliton solutions:
\begin{equation}
l_j=\bar{k}_j,\quad e^{\delta_j}=\rho e^{\bar{\alpha}_j},\quad j=1, 2.
\end{equation}
Hence two-soliton solution of the local reduced CLL equation (\ref{redlocalCLL}) is
\begin{equation}\label{2SSlocalCLL}\displaystyle
q(x,t)=\frac{ e^{\theta_1}+e^{\theta_2}+\rho\tilde{\gamma}_1e^{\theta_1+\theta_2+\bar{\theta}_1}+\rho\tilde{\gamma}_2e^{\theta_1+\theta_2+\bar{\theta}_2}}
{1-\frac{i\rho}{2}\sum_{1\leq m,n\leq 2} \frac{k_m}{(k_m+\bar{k}_n)^2}e^{\theta_m+\bar{\theta}_n}+\rho^2\tilde{M}_1e^{\theta_1+\theta_2+\bar{\theta}_1+\bar{\theta}_2}        },
\end{equation}
and two-soliton solution of the local reduced GI equation (\ref{redlocalGI}) is
\begin{equation}\label{2SSlocalGI}\displaystyle
q(x,t)=\frac{e^{\theta_1}+e^{\theta_2}+\rho\tilde{\gamma}_1e^{\theta_1+\theta_2+\bar{\theta}_1}+\rho\tilde{\gamma}_2e^{\theta_1+\theta_2+\bar{\theta}_2}}
{1+\frac{i\rho}{2}\sum_{1\leq m,n\leq 2} \frac{\bar{k}_n}{(k_m+\bar{k}_n)^2}e^{\theta_m+\bar{\theta}_n}+\rho^2\tilde{M}_2e^{\theta_1+\theta_2+\bar{\theta}_1+\bar{\theta}_2}        },
\end{equation}
where $\theta_j=k_jx+\frac{i}{a}k_j^2t+\alpha_j$, $j=1, 2$. Here $\tilde{\gamma}_p=\gamma_p\Big|_{l_j=\bar{k}_j}$, $\tilde{M}_p=M_p\Big|_{l_j=\bar{k}_j}$ for $p,j=1,2$, where $\gamma_j$, $j=1,2$ are given in (\ref{gammabeta}), $M_1$ and $M_2$ are given in (\ref{M_1M_2}).

\subsubsection{Two-soliton solution of the local reduced KN equation}

\noindent A.(i)\, $r(x,t)=\rho \bar{q}(x,t), a=\bar{a}$, $\rho\in \mathbb{R}$.

In this case, Type 1 yields the constraints
\begin{equation}
l_j=\bar{k}_j,\quad e^{\delta_j}=\rho e^{\bar{\alpha}_j},\quad j=1, 2.
\end{equation}
Hence two-soliton solution of the local reduced KN equation (\ref{redlocalKN}) is
{\small\begin{equation}\label{2SSlocalKN}\displaystyle
q(x,t)=\frac{[e^{\theta_1}+e^{\theta_2}+\rho\tilde{\gamma}_1e^{\theta_1+\theta_2+\bar{\theta}_1}
+\rho\tilde{\gamma}_2e^{\theta_1+\theta_2+\bar{\theta}_2}][1+\frac{i\rho}{2}\sum_{1\leq m,n\leq 2}\frac{\bar{k}_n}{(k_m+\bar{k}_n)}e^{\theta_m+\bar{\theta}_n}+\rho^2\tilde{M}_2e^{\theta_1+\theta_2+\bar{\theta}_1+\bar{\theta}_2} ]}
{[1-\frac{i\rho}{2}\sum_{1\leq m,n\leq 2}\frac{k_m}{(k_m+\bar{k}_n)}e^{\theta_m+\bar{\theta}_n}+\rho^2\tilde{M}_1e^{\theta_1+\theta_2+\bar{\theta}_1+\bar{\theta}_2}]^2},
\end{equation}}
where $\theta_j=k_jx+\frac{i}{a}k_j^2t+\alpha_j$, $j=1, 2$. Here $\tilde{\gamma}_p=\gamma_p\Big|_{l_j=\bar{k}_j}$, $\tilde{M}_p=M_p\Big|_{l_j=\bar{k}_j}$ for $p,j=1,2$, where $\gamma_j$, $j=1,2$ are given in (\ref{gammabeta}), $M_1$ and $M_2$ are given in (\ref{M_1M_2}).\\

\begin{rem}\label{rem1}
 If $k_1$ or $k_2$ is a pure imaginary number then all the solutions (\ref{2SSlocalCLL}), (\ref{2SSlocalGI}), and (\ref{2SSlocalKN}) become trivial solutions $q(x,t)=0$. Furthermore, the real-valued two-soliton solutions $|q(x,t)|^2$ obtained from (\ref{2SSlocalCLL}), (\ref{2SSlocalGI}), and (\ref{2SSlocalKN}) are all same for the local reduced CLL, GI, and KN equations.
\end{rem}

\noindent \textbf{Example 6.} Here let us consider two set of parameters; $(k_1=1, k_2=2, a=2, e^{\alpha_1}=-1, e^{\alpha_2}=\rho=1)$ and
$(k_1=1-i, k_2=1+i, a=2, e^{\alpha_1}=e^{\alpha_2}=1, \rho=-1)$. For the local reduced CLL, GI, and KN equations, under the first set of parameters we get breather type two-soliton solution $|q(x,t)|^2$ given in Figure 6(a) and the second set of parameters gives the bell type two-soliton solution presented in Figure 6(b).
\begin{center}
\begin{figure}[h!]
\centering
\subfloat[]{\includegraphics[width=0.34\textwidth]{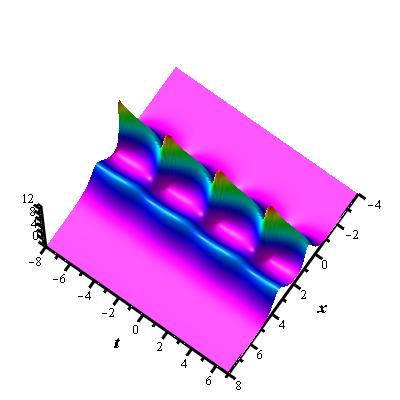}}\hspace{2cm}
\subfloat[] {\includegraphics[width=0.34\textwidth]{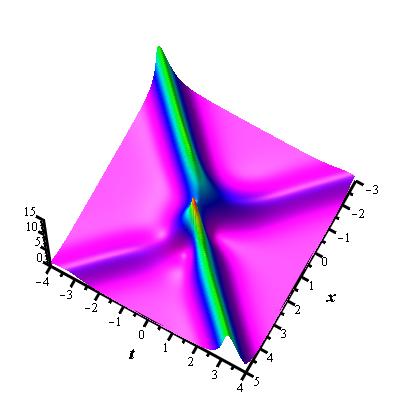}}
\caption{ (a) Breather type two-soliton solution with $k_1=1, k_2=2, a=2, e^{\alpha_1}=-1, e^{\alpha_2}=\rho=1$, (b) Bell type two-soliton solution with $k_1=1-i, k_2=1+i, a=2, e^{\alpha_1}=e^{\alpha_2}=1, \rho=-1$.}
\end{figure}
\end{center}
\squeezeup

\subsection{Two-soliton solutions of the shifted nonlocal reduced equations}

Similar to one-soliton solutions we find two-soliton solutions of the shifted nonlocal reduced CLL, GI, and KN
equations by using Type 1 and Type 2 approaches with the reduction formulas and two-soliton solutions of
CLL, GI, and KN systems. Note that since the expressions for two-soliton solutions are very long, we will
not give them explicitly. Here we will only present the constraints to be satisfied by the parameters of two-soliton solutions
of the reduced equations and some particular examples.

\subsubsection{Two-soliton solutions of the shifted nonlocal reduced CLL and GI equations}

\noindent B.(i)\, $r(x,t)=\rho q(-x+x_0,-t+t_0)$, $\rho, x_0, t_0 \in \mathbb{R}$.

Since Type 1 gives trivial solution, we use Type 2 and get the following constraints on the parameters of two-soliton solutions:
\begin{equation}\displaystyle
e^{\delta_1}=\sigma_1\sqrt{\frac{-2i\rho l_2}{k_1k_2}}\frac{(k_1+l_1)(k_2+l_1)}{(l_1-l_2)}e^{\frac{-l_1x_0-m_1t_0}{2}},\,
e^{\delta_2}=\sigma_2\sqrt{\frac{-2i\rho l_1}{k_1k_2}}\frac{(k_1+l_2)(k_2+l_2)}{(l_1-l_2)}e^{\frac{-l_2x_0-m_2t_0}{2}},
\end{equation}
and
\begin{equation}\displaystyle
e^{\alpha_1}=\sigma_3\sqrt{\frac{2ik_2}{\rho l_1l_2}}\frac{(k_1+l_2)(k_1+l_1)}{(k_1-k_2)}e^{\frac{-k_1x_0-\omega_1t_0}{2}},\,
e^{\alpha_2}=\sigma_4\sqrt{\frac{2ik_1}{\rho l_1l_2}}\frac{(k_2+l_1)(k_2+l_2)}{(k_1-k_2)}e^{\frac{-k_2x_0-\omega_2t_0}{2}},
\end{equation}
for the real shifted nonlocal space-time reversal reduced CLL equation (\ref{redrealSTrevCLL}), and
\begin{equation}\displaystyle
e^{\delta_1}=\sigma_1\sqrt{\frac{-2i\rho}{l_1}}\frac{(k_1+l_1)(k_2+l_1)}{(l_1-l_2)}e^{\frac{-l_1x_0-m_1t_0}{2}},\,
e^{\delta_2}=\sigma_2\sqrt{\frac{-2i\rho}{l_2}}\frac{(k_1+l_2)(k_2+l_2)}{(l_1-l_2)}e^{\frac{-l_2x_0-m_2t_0}{2}},
\end{equation}
and
\begin{equation}\displaystyle
e^{\alpha_1}=\sigma_3\sqrt{\frac{2i}{\rho k_1}}\frac{(k_1+l_2)(k_1+l_1)}{(k_1-k_2)}e^{\frac{-k_1x_0-\omega_1t_0}{2}},\,
e^{\alpha_2}=\sigma_4\sqrt{\frac{2i}{\rho k_2}}\frac{(k_2+l_1)(k_2+l_2)}{(k_1-k_2)}e^{\frac{-k_2x_0-\omega_2t_0}{2}},
\end{equation}
for the real shifted nonlocal space-time reversal reduced GI equation (\ref{redrealSTrevGI}). Here $\sigma_j=\pm 1$, $j=1,2,3,4$.

\noindent \textbf{Example 7.} Let $k_1=l_1=1, k_2=l_2=\frac{1}{2}, a=1, \sigma_j=\rho=x_0=1$ for $j=1,2,3,4$ and $t_0=-2$ in two-soliton solution for
the real shifted nonlocal space-time reversal reduced CLL equation (\ref{redrealSTrevCLL}). We get breather type
two-soliton solution given in Figure 7.
\begin{center}
\begin{figure}[h]
\centering
\begin{minipage}[t]{1\linewidth}
\centering
\includegraphics[angle=0,scale=.36]{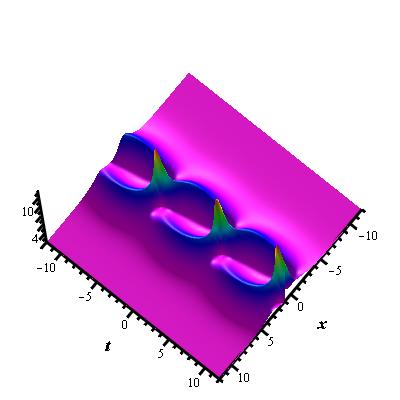}
\caption{Breather type
two-soliton solution with the parameters $k_1=l_1=1, k_2=l_2=\frac{1}{2}, a=1, \rho=x_0=\sigma_j=1$ for $j=1,2,3,4$ and $t_0=-2$ for the real shifted nonlocal space-time reversal reduced CLL equation.}
\end{minipage}%
\end{figure}
\end{center}
\squeezeup

\noindent \textbf{Example 8.} Now take $k_1=l_1=1, k_2=l_2=\frac{1}{2}, a=i, \sigma_2=\sigma_3=\sigma_4=\rho=x_0=1, \sigma_1=-1, t_0=-2$
in two-soliton solution for
the real shifted nonlocal space-time reversal reduced GI equation (\ref{redrealSTrevGI}). We get two asymptotically decaying waves illustrated in Figure 8.
\begin{center}
\begin{figure}[h]
\centering
\begin{minipage}[t]{1\linewidth}
\centering
\includegraphics[angle=0,scale=.35]{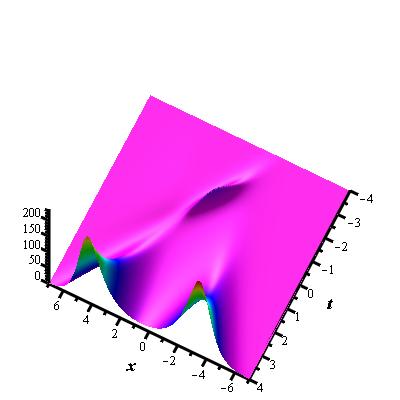}
\caption{Two asymptotically decaying waves with the parameters $k_1=l_1=1, k_2=l_2=\frac{1}{2}, a=i, \sigma_2=\sigma_3=\sigma_4=\rho=x_0=1, \sigma_1=-1, t_0=-2$ for the real shifted nonlocal space-time reversal reduced GI equation.}
\end{minipage}%
\end{figure}
\end{center}
\squeezeup

\noindent B.(ii)\, $r(x,t)=\rho \bar{q}(x,-t+t_0)$, $\rho, t_0 \in \mathbb{R}$, $a=-\bar{a}$.

When we use Type 1, we get the following constraints that must be satisfied by the parameters of two-soliton solutions
of both complex shifted nonlocal time reversal reduced CLL and GI equations:
\begin{equation}
l_j=\bar{k}_j,\quad e^{\delta_j}=\rho e^{\bar{\omega}_jt_0+\bar{\alpha}_j},\quad j=1,2,
\end{equation}
yielding $m_j=-\bar{\omega}_j$, $j=1, 2$.

\subsubsection{Two-soliton solutions of the shifted nonlocal reduced KN equations}

\noindent B.(i)\, $r(x,t)=\rho q(-x+x_0,-t+t_0)$, $\rho, x_0, t_0 \in \mathbb{R}$.

Here Type 1 yields trivial solution. Applying Type 2 to the reduction formula with two-soliton solution of KN system yields
the constraints
\begin{equation}\displaystyle
e^{\delta_1}=\sigma_1\sqrt{-2i\rho l_1}\frac{l_2(k_1+l_1)(k_2+l_1)}{k_1k_2(l_1-l_2)}e^{\frac{-l_1x_0-m_1t_0}{2}},\,
e^{\delta_2}=\sigma_2\sqrt{-2i\rho l_2}\frac{l_1(k_1+l_2)(k_2+l_2)}{k_1k_2(l_1-l_2)}e^{\frac{-l_2x_0-m_2t_0}{2}},
\end{equation}
and
\begin{equation}\displaystyle
e^{\alpha_1}=\sigma_3\sqrt{\frac{2ik_1}{\rho}}\frac{k_2(k_1+l_2)(k_1+l_1)}{l_1l_2(k_1-k_2)}e^{\frac{-k_1x_0-\omega_1t_0}{2}},\,
e^{\alpha_2}=\sigma_4\sqrt{\frac{2ik_2}{\rho}}\frac{k_1(k_2+l_1)(k_2+l_2)}{l_1l_2(k_1-k_2)}e^{\frac{-k_2x_0-\omega_2t_0}{2}},
\end{equation}
for the real shifted nonlocal space-time reversal reduced KN equation (\ref{redrealSTrevKN}). Here $\sigma_j=\pm 1$, $j=1,2,3,4$.\\

\noindent \textbf{Example 9.} Choose $k_1=\frac{1}{2}, k_2=\frac{1}{8}, l_1=2, l_2=1, a=2i, \sigma_j=x_0=\rho=1$ for $j=1,2,3,4$ and $t_0=-2$
in two-soliton solution for
the equation (\ref{redrealSTrevKN}). We obtain two asymptotically decaying waves illustrated in Figure 9.
\begin{center}
\begin{figure}[h]
\centering
\begin{minipage}[t]{1\linewidth}
\centering
\includegraphics[angle=0,scale=.35]{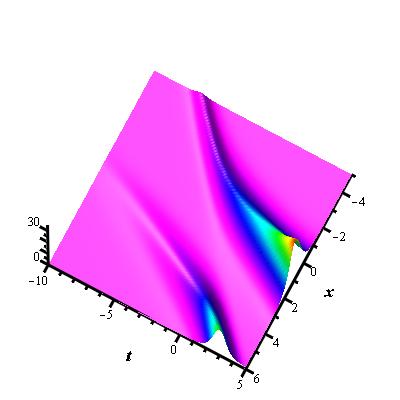}
\caption{Two asymptotically decaying waves with the parameters $k_1=\frac{1}{2}, k_2=\frac{1}{8}, l_1=2, l_2=1, a=2i, x_0=\rho=\sigma_j=1$ where $j=1,2,3,4$ and $t_0=-2$ for the real shifted nonlocal space-time reversal reduced KN equation.}
\end{minipage}%
\end{figure}
\end{center}
\squeezeup

\noindent B.(ii)\, $r(x,t)=\rho \bar{q}(x,-t+t_0)$, $\rho, t_0 \in \mathbb{R}$, $a=-\bar{a}$.

In this case, Type 1 approach yields the following constraints that must be satisfied by the parameters of two-soliton solutions
of the complex shifted nonlocal time reversal reduced KN equation:
\begin{equation}
l_j=\bar{k}_j,\quad e^{\delta_j}=\rho e^{\bar{\omega}_jt_0+\bar{\alpha}_j},\quad j=1, 2,
\end{equation}
yielding $m_j=-\bar{\omega}_j$, $j=1, 2$.\\

\begin{rem}\label{rem2}
  Let $k_1$ or $k_2$ be a pure imaginary number. In this case, two-soliton solutions of the complex shifted nonlocal time reversal reduced CLL, GI, and KN equations become same and indeed trivial. Moreover, the real-valued solutions $|q(x,t)|^2$ obtained from two-soliton solutions of the complex shifted nonlocal time reversal reduced CLL, GI, and KN equations are same.
\end{rem}

\noindent \textbf{Example 10.} Take the parameters of two-soliton solutions of the complex shifted nonlocal time reversal reduced CLL, GI, and KN equations as $k_1=\frac{1}{2}, k_2=-\frac{1}{10}, a=\frac{i}{3}, e^{\alpha_1}=x_0=\rho=1, e^{\alpha_2}=-1, t_0=-2$. Then we get the same solution $|q(x,t)|^2$ for these equations. We have two asymptotically decaying waves presented in Figure 11.
\begin{center}
\begin{figure}[h]
\centering
\begin{minipage}[t]{1\linewidth}
\centering
\includegraphics[angle=0,scale=.35]{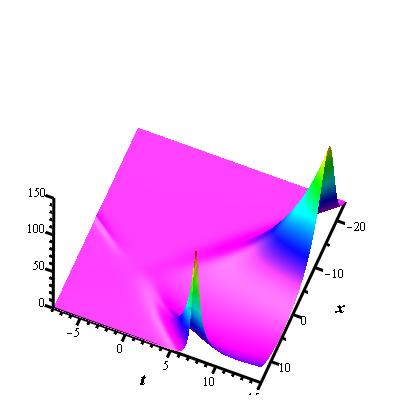}
\caption{Two asymptotically decaying waves with the parameters $k_1=\frac{1}{2}, k_2=-\frac{1}{10}, a=\frac{i}{3}, e^{\alpha_1}=x_0=\rho=1, e^{\alpha_2}=-1, t_0=-2$ for the complex shifted nonlocal time reversal reduced CLL, GI, and KN equations.}
\end{minipage}%
\end{figure}
\end{center}
\squeezeup

\section{Conclusion}
In this paper we considered the integrable Kundu type system including a free parameter $\beta$. We obtained
all possible local and shifted nonlocal consistent reductions of this system. We used two different bilinearizing transformations on this system
and obtained Hirota bilinear forms of the famous systems corresponding to three particular values of $\beta$; CLL system ($\beta=-\frac{1}{4}$), GI system ($\beta=0$), and KN system ($\beta=-\frac{1}{2}$). Via the Hirota method we obtained one- and two-soliton solutions of these coupled systems.
By using these solutions with the reduction formulas we also obtained one- and two-soliton solutions of the local and shifted nonlocal real space-time reversal and complex time reversal reduced CLL, GI, and KN equations.

We observed that one- and two-soliton solutions $|q(x,t)|^2$ of the reduced local and shifted nonlocal complex time-reversal CLL, GI, and KN equations obtained by Type 1 are same. But the solutions obtained by Type 2 for the shifted nonlocal real space-time reversal reduced CLL, GI, and KN equations are different. One-soliton solutions $q(x,t)$ of these equations are also related to the one-soliton solutions $u(x,t)$ of the standard unshifted nonlocal real space-time reversal equations by the relation $q(x,t)=u(x-\frac{x_0}{2},t-\frac{t_0}{2})$.

One can also find $N$-soliton solutions of these equations via Hirota method by using the expansion (\ref{generalexpansion}) with (\ref{generalg_1}) and (\ref{generalh_1}) for any $N$. Using different methods like inverse scattering transformation and Darboux transformation can give different solutions for the shifted nonlocal CLL, GI, and KN equations. We also remark that since the shifted nonlocal CLL, GI, and KN equations reduce to the standard unshifted nonlocal CLL, GI, and KN equations, respectively, by letting $x_0=t_0=0$, solutions of the shifted nonlocal equations also reduce to solutions of the standard nonlocal equations as $x_0=t_0=0$.



\begin{thebibliography}{00}

\bibitem{Kakei} S. Kakei, N. Sasa, and J. Satsuma, Bilinearization of a generalized derivative nonlinear Schr\"{o}dinger equation, J. Phys. Soc.
Japan \textbf{64} (5), 1519--1523, 1995.

\bibitem{Fan} E.G. Fan, A family of completely integrable multi-Hamiltonian systems explicitly related to some celebrated equations,
J. Math. Phys. \textbf{42} (9), 4327--4344, 2001.

\bibitem{HonFan} Y.C. Hon, E.G. Fan, Uniformly constructing finite-band solutions for a family of derivative nonlinear Schr\"{o}dinger
equations, Chaos, Solitons Fractals \textbf{24}, 1087-–1096, 2005.

\bibitem{Abhinav} K. Abhinav, P. Guha, and I. Mukherjee, Study of quasi-integrable and nonholonomic deformation of equations in the NLS and DNLS
hierarchy, J. Math. Phys. \textbf{59}, 101507, 2018.

\bibitem{TD} M. Tao, H. Dong, N-soliton solutions of the coupled Kundu equations based on the Riemann-Hilbert method, Math. Probl. Eng. \textbf{2019},
Art. ID 3085367, 2019.

\bibitem{Kundu1} A. Kundu, Landau-Lifshitz and higher-order nonlinear systems gauge generated from nonlinear Schr\"{o}dinger-type equations,
J. Math. Phys. \textbf{25}, 3433--3438, 1984.

\bibitem{Kundu2} A. Kundu, Exact solutions to higher-order nonlinear equations through gauge transformation, Physica D \textbf{25}, 399--406, 1987.

\bibitem{Clarkson} P.A. Clarkson, C.M. Cosgrove, Painlev\'{e} analysis of the non-linear Schr\"{o}dinger family of equations, J. Phys. A \textbf{20}, 2003, 1987.

\bibitem{WenZFan} L. Wen, N. Zhang, and E. Fan, N-soliton solution of the Kundu-type equation via Riemann-Hilbert approach, Acta Math. Sci. Ser. B \textbf{40} (1), 113--126, 2020.

\bibitem{CLL1} H.H. Chen, Y.C. Lee, and C.S. Liu, Integraility of nonlinear Hamiltonian systems by inverse scattering method, Phys. Scr.
\textbf{20}, 490--492, 1979.

\bibitem{CLL2} A. Nakamura, H.H. Chen, Multi-soliton solutions of a derivative nonlinear Schr\"{o}dinger equation, J. Phys. Soc. Japan
\textbf{49} (2), 1980.

\bibitem{CLL3} B. Yang, W.G. Zhang, H.Q. Zhang, and S.B. Pei, Generalized Darboux transformation and rational soliton
solutions for Chen-Lee-Liu equation, Appl. Math. Comput. \textbf{242}, 863-–876, 2014.

\bibitem{CLL4} N. Zhang, T.C. Xia, and E.G. Fan, A Riemann-Hilbert approach to the Chen-Lee-Liu equation on the half line,
Acta Math. Appl. Sin. Engl. Ser. \textbf{34} (3), 493-–515, 2018.

\bibitem{CLL5} Y.S. Zhang, L.J. Guo, J.S. He, and Z.X. Zhou, Darboux transformation of the second-type
derivative nonlinear Schr\"{o}dinger equation, Lett. Math. Phys. \textbf{105} (6), 853-–891, 2015.

\bibitem{GI1} V.S. Gerdjikov, M.I. Ivanov, The quadratic bundle of general
form and the nonlinear evolution equations; I. Hierarchies
of Hamiltonian structures, Bulg. J. Phys. \textbf{10}, 13, 1983.

\bibitem{GI2} V.S. Gerdjikov, M.I. Ivanov, The quadratic bundle of general
form and the nonlinear evolution equations; II. Hierarchies
of Hamiltonian structures, Bulg. J. Phys. \textbf{10}, 130, 1983.

\bibitem{GI3} Yu.N. Sidorenko, K. Prikarpatskii, Periodic problem for nonlinear Ablowitz model, J. Math. Sci. \textbf{65} (6), 1921--1927, 1993.

\bibitem{GI4} E.G. Fan, Integrable evolution systems based on Gerdjikov-Ivanov equations, bi-Hamiltonian structure, finite-dimensional
integrable systems and N-fold Darboux transformation, J. Math. Phys. \textbf{41}, 7769–-7782, 2000.

\bibitem{GI5} E.G. Fan, Darboux transformation and soliton-like solutions for the
Gerdjikov–Ivanov equation, J. Phys. A: Math. Gen. \textbf{33}, 6925–6933, 2000.

\bibitem{GI6} Y. Hou, E.G. Fan, and P. Zhao, Algebro-geometric solutions for the Gerdjikov-Ivanov hierarchy, J. Math. Phys. \textbf{54}, 073505, 2013.

\bibitem{GI7} H. Yilmaz, Exact solutions of the Gerdjikov-Ivanov equation
using Darboux transformations, J. Nonlinear Math. Phys. \textbf{22} (1), 32--46, 2015.

\bibitem{KN1} D.J. Kaup, A.C. Newell, An exact solution for a derivative nonlinear Schr\"{o}dinger equation, J. Math. Phys. \textbf{19}, 798, 1978.

\bibitem{KN2} G. Tu, The trace identity, a powerful tool for constructing the Hamiltonian structure of
integrable systems, J. Math. Phys. \textbf{30}, 330--338, 1989.

\bibitem{KN3} S.W. Xu, J.S. He, and L.H. Wang, The Darboux transformation of the derivative nonlinear Schr\"{o}dinger equation,
J. Phys. A: Math. Theor. \textbf{44} (30), 305203, 2011.

\bibitem{KN4} B. Guo, L. Ling, and Q.P. Liu, High-order solutions and generalized Darboux transformations of derivative nonlinear Schr\"{o}dinger equations, Stud. Appl. Math. \textbf{130}, 317--344, 2012.

\bibitem{KN5} Y.S. Zhang, L.J. Guo, S.W. Xu, Z.W. Wu, and J.S. He, The hierarchy of higher order solutions of the derivative
nonlinear Schr\"{o}dinger equation, Commun. Nonlinear Sci. Numer. Simulat. \textbf{19} (6), 1706–-1722, 2014

\bibitem{KN6} W. Liu, Y. Zhang, and J. He, Rogue wave on a periodic background for Kaup-Newell equation, Rom. Rep. Phys. \textbf{70}, 106, 2018.



\bibitem{Wadati} M. Wadati, K. Sogo, Gauge transformations in soliton theory, J. Phys. Soc. Japan \textbf{52}, 394, 1983.



    \bibitem{AbMu1} M.J. Ablowitz, Z.H. Musslimani, Integrable nonlocal nonlinear Schr\"{o}dinger equation, Phys. Rev. Lett. \textbf{110}, 064105, 2013.

\bibitem{AbMu2} M.J. Ablowitz, Z.H. Musslimani, Inverse scattering transform for the integrable nonlocal nonlinear Schr\"{o}dinger equation, Nonlinearity \textbf{29}, 915--946, 2016.

\bibitem{AbMu3} M.J. Ablowitz, Z.H. Musslimani, Integrable nonlocal nonlinear equations, Stud. App. Math. \textbf{139} (1), 7--59, 2016.

\bibitem{origin} M. G\"{u}rses, A. Pekcan, and K. Zheltukhin, Discrete symmetries and nonlocal reductions, Phys. Lett. A \textbf{384}, 120065, 2020.

\bibitem{chen} K. Chen, X. Deng, S. Lou, and D. Zhang, Solutions of local and nonlocal equations reduced from the AKNS hierarchy, Stud. Appl. Math. \textbf{141} (1), 113--141, 2018.

\bibitem{FLAH} B.F. Feng, X.D. Luo, M.J. Ablowitz, and Z.H. Musslimani, General soliton solution to a nonlocal nonlinear Schr\"{o}dinger
equation with zero and nonzero boundary conditions, Nonlinearity \textbf{31} (12), 5385--5409, 2018.


\bibitem{gerd} V.S. Gerdjikov, A. Saxena, Complete integrability of nonlocal nonlinear Schr\"{o}dinger equation, J. Math. Phys. \textbf{58} (1), 013502, 2017.



\bibitem{GurPek1}  M. G\"{u}rses, A. Pekcan, Nonlocal nonlinear Schr\"{o}dinger equations and their soliton solutions, J. Math. Phys. \textbf{59}, 051501, 2018.

\bibitem{GurPek3} M. G\"{u}rses, A. Pekcan, Integrable nonlocal reductions, Symmetries, Differential Equations and Applications SDEA-III, Istanbul, Turkey, August 2017, in: V.G. Kac, P.J. Olver, P. Winternitz, T. Ozer (Eds), Springer Proceedings in Mathematics and Statistics, \textbf{266}, 27--52, 2018.


\bibitem{huang} X. Huang, L. Ling, Soliton solutions for the nonlocal nonlinear Schr\"{o}dinger equation, Eur. Phys. J. Plus \textbf{131}, 148, 2016.


\bibitem{li}  M. Li, T. Xu, Dark and antidark soliton interactions in the nonlocal nonlinear Schr\"{o}dinger equation
with the self-induced parity-time-symmetric potential, Phys. Rev. E \textbf{91}, 033202, 2015.


\bibitem{aflm1} M.J. Ablowitz, B.F. Feng, X.D. Luo, and Z.H. Musslimani, Inverse scattering transform for the nonlocal reverse space-time
nonlinear Schr\"{o}dinger equation, Theor. Math. Phys. \textbf{196} (3), 1241--1267, 2018.


\bibitem{Wen} X.Y. Wen, Z. Yan, and Y. Yang, Dynamics of higher-order rational solitons for the nonlocal nonlinear Schr\"{o}dinger
equation with the self-induced parity-time-symmetric potential, Chaos \textbf{26}, 063123, 2015.

\bibitem{Sax} A. Khare, A. Saxena, Periodic and hyperbolic soliton solutions of a number of nonlocal nonlinear
equations, J. Math. Phys. \textbf{56}, 032104, 2015.

\bibitem{jianke} J. Yang, General N-solitons and their dynamics in several nonlocal nonlinear Schr\"{o}dinger equations, Phys. Lett A \textbf{383} (4), 328--337, 2019.

\bibitem{XLLLZ} T. Xu, S. Lan, M. Li, L.L. Li, and G.W. Zhang, Mixed soliton solutions of the defocusing nonlocal nonlinear Schr\"{o}dinger equation, Phys. D
\textbf{390}, 47--61, 2019.

\bibitem{XCLM} T. Xu, Y. Chen, M. Li, and D.X. Meng, General stationary solutions of the nonlocal nonlinear Schr\"{o}dinger equation and their relevance
to the $\mathcal{PT}$-symmetric system, Chaos \textbf{29}, 123124, 2019.

\bibitem{Ma1} W.X. Ma, Inverse scattering for nonlocal reverse-time nonlinear Schr\"{o}dinger equations, Appl. Math. Lett. \textbf{102}, 106161, 2020.


\bibitem{GurPek2}  M. G{\" u}rses,  A. Pekcan, Nonlocal nonlinear modified KdV equations and their soliton solutions, Commun. Nonlinear Sci. Numer.
Simul. \textbf{67}, 427, 2019.

\bibitem{JZ1} J.L. Ji, Z.N. Zhu, On a nonlocal modified Korteweg-de Vries equation: Integrability, Darboux transformation and soliton solutions,
Commun. Non. Sci. Numer. Simul. \textbf{42}, 699, 2017.

\bibitem{JZ2} J.L. Ji, Z.N. Zhu, Soliton solutions of an integrable nonlocal modified Korteweg-de Vries equation through
inverse scattering transform, J. Math. Anal. Appl. \textbf{453}, 973, 2017.

\bibitem{ma}   L.Y. Ma, S.F. Shen, and Z.N. Zhu, Soliton solution and gauge equivalence for an integrable nonlocal complex modified Korteweg-de Vries
equation, J. Math. Phys. \textbf{58}, 103501, 2017.

\bibitem{Shi} X. Shi, P. Lv, and C. Qi, Explicit solutions to a nonlocal 2-component complex modified Korteweg-de Vries equation, Appl. Math. Lett.
\textbf{100}, 106043, 2020.

\bibitem{Luo} X.D. Luo, Inverse scattering transform for the complex reverse space-time nonlocal modified Korteweg-de Vries equation with nonzero boundary conditions and constant phase shift, Chaos \textbf{29}, 073118, 2019.

\bibitem{Yan} G. Zhang, Z. Yan, Inverse scattering transforms and soliton solutions of focusing and defocusing nonlocal mKdV equations with nonzero boundary conditions, Phys. D \textbf{402}, 132170, 2020.

\bibitem{LZYY} M. Li, Y. Zhang, R. Ye, and Y. Lou, Exact solutions of the nonlocal Gerdjikov-Ivanov equation, Commun. Theor. Phys.
\textbf{73} (10), 2021.

\bibitem{ZD} Y. Zhang, H.H. Dong, N-soliton solutions to the multi-component nonlocal Gerdjikov-Ivanov equation via Riemann-Hilbert problem with zero
boundary conditions, Appl. Math. Lett. \textbf{125}, 107770, 2022.


\bibitem{Gurses} M. G\"{u}rses, Nonlocal Fordy-Kulish equations on symmetric spaces, Phys. Lett. A \textbf{381}, 1791--1794, 2017.

\bibitem{GurPekKos} M. G\"{u}rses, A. Pekcan, and K. Zheltukhin, Nonlocal hydrodynamic type of equations, Commun. Nonlinear Sci. Numer. Simul. \textbf{85}, 105242, 2020.

\bibitem{GurPek6} M. G\"{u}rses, A. Pekcan, Nonlocal KdV equations, Phys. Lett. A \textbf{384} (35), 126894, 2020.



\bibitem{RZFH} J. Rao, Y. Zhang, A.S. Fokas, and J. He, Rogue waves of the nonlocal Davey-Stewartson I equation, Nonlinearity \textbf{31}, 4090--4107, 2018.

\bibitem{XLHCY} T. Xu, M. Li, Y. Huang, Y. Chen, and C. Yu, Nonsingular localized wave solutions for the nonlocal Davey-Stewartson I equation
with zero background, Modern Phys. Lett. B \textbf{31} (35), 1750338, 2017.

\bibitem{ZXZhou} Z.X. Zhou, Darboux transformations global explicit solutions for nonlocal Davey-Stewartson I equation, Stud. Appl. Math. \textbf{141} (2) 186--204, 2018.


\bibitem{ZL} Y. Zhang, Y. Liu, Breather and lump solutions for nonlocal Davey-Stewartson II equation, Nonlinear Dyn. \textbf{96}, 107--113, 2019.

\bibitem{Rao} J. Rao, Y. Cheng, and J. He, Rational and semirational solutions of the nonlocal Davey-Stewartson equations, Stud. Appl. Math. \textbf{139} (4), 2017.



\bibitem{GurPek4}  M. G\"{u}rses, A. Pekcan, (2+1)-dimensional local and nonlocal reductions of the negative AKNS system: Soliton solutions, Commun. Nonlinear Sci. Numer. Simul. \textbf{71}, 161--173, 2019.

\bibitem{GurPek5} M. G\"{u}rses, A. Pekcan, (2+1)-dimensional AKNS($-N$) systems II, Commun. Nonlinear Sci. Numer.
Simul. \textbf{97}, 105736, 2021.




\bibitem{AbMu4} M.J. Ablowitz, Z.H. Musslimani, Integrable space-time shifted nonlocal nonlinear equations, Phys. Lett. A \textbf{409}, 127516, 2021.

\bibitem{GurPek7}  M. G\"{u}rses, A. Pekcan, Soliton solutions of the shifted nonlocal NLS and MKdV equations, arXiv:2106.14252v2 [nlin.SI].

\bibitem{Dajun} S.M. Liu, J. Wang, and D.J. Zhang, Solutions to integrable space-time shifted nonlocal equations, arXiv:2107.04183v1 [nlin.SI].

\bibitem{ZGC} J.B. Zhang, Y.Y. Gongye, and S.T. Chen, Soliton solutions to the coupled Gerdjikov-Ivanov equation with rogue-wave-like phenomena, Chin. Phys. Lett. \textbf{34} (9), 090201, 2017.


\bibitem{Hirota1} R. Hirota, The Direct Method in Soliton Theory, Cambridge University Press, Cambridge, 2004.


\bibitem{YangYang} B. Yang, J. Yang, Transformations between nonlocal and local integrable equations, Stud. Appl. Math.
\textbf{140} (2), 178--201, 2017.

\bibitem{MuGu} I. Mukherjee, P. Guha, A study of nonholonomic deformations of nonlocal integrable systems
belonging to the nonlinear Schr\"{o}dinger family, Rus. J. Nonlin. Dyn. \textbf{15} (3), 293--307, 2019.

\bibitem{Valchev} T. Valchev, On Mikhailov's reduction group, Phys. Lett. A \textbf{379}, 1877--1880, 2015.

\bibitem{Zhou} Z.X. Zhou, Darboux transformations and global solutions for a nonlocal derivative nonlinear Schr\"{o}dinger
equation, Commun. Nonlinear Sci. Numer. Simulat. \textbf{62}, 480--488, 2018.




\end{thebibliography}
\end{document}